\documentclass[universe,review,accept,pdftex,moreauthors]{Definitions/mdpi} 

\firstpage{1} 
\pubvolume{1}
\issuenum{1}
\articlenumber{0}
\pubyear{2023}
\copyrightyear{2023}
\datereceived{ } 
\daterevised{ } 
\dateaccepted{ } 
\datepublished{ } 
\hreflink{https://doi.org/} 

\DeclareRobustCommand{\ion}[2]{%
\relax\ifmmode
\ifx\testbx\f@series
{\mathbf{#1\,\mathsc{#2}}}\else
{\mathrm{#1\,\mathsc{#2}}}\fi
\else\textup{#1\,{\mdseries\textsc{#2}}}%
\fi}

\def\rfe{R$_\mathrm{FeII}$}
\def\hb{H$\beta$}
\def\mbh{$M_{\rm BH}$\/}
\def\cc{cm$^{-3}$}
\def\feii{{Fe {\sc ii}}}
\def\kms{km~s$^{-1}$}

\def\msun{M$_\odot$\/}

\def\rblr{R$_{\rm BLR}$}
\def\lopt{L$_{\rm 5100}$}
\def\lbol{L$_{\rm bol}$}
\def\ledd{L$_{\rm Edd}$}
\def\lledd{L$_{\rm bol}$/L$_{\rm Edd}$}
\def\mdot{$\dot{\mathcal{M}}$}

\usepackage{rotating}

\Title{Spectral variability studies in Active Galactic Nuclei: Exploring continuum and emission line regions in the age of LSST and JWST}

\TitleCitation{Selected studies in spectral line variability in AGNs}

\Author{Swayamtrupta Panda$^{1,\dagger}$\orcidA{}, Paola Marziani$^{2}$\orcidB{}, Bo\.zena Czerny$^{3}$\orcidC{}, Alberto Rodr\'iguez-Ardila$^{1,4}$\orcidD{}, Francisco Pozo Nu\~nez$^{5}$\orcidE{}}

\AuthorNames{Swayamtrupta Panda, Paola Marziani, Bo\.zena Czerny, Alberto Rodr\'iguez Ardila, Francisco Pozo Nu\~nez}

\AuthorCitation{Panda, S.; Marziani, P.; Czerny, B.; Rodr\'guez Ardila, A.; Pozo Nu\~nez, F.; }

\address{%
$^{1}$ \quad Laborat\'orio Nacional de Astrofísica, Rua dos Estados Unidos, 154, Bairro nas Nações, Itajubá, MG, Brazil\\
$^{2}$ \quad INAF-Astronomical Observatory of Padova, Vicolo dell'Osservatorio, 5, 35122 Padova PD, Italy\\
$^{3}$ \quad Center for Theoretical Physics, Polish Academy of Sciences, Al. Lotnik\'ow 32/46, 02-668 Warsaw, Poland\\
$^{4}$ \quad Divis\~ao de Astrof\'isica, Instituto Nacional de Pesquisas Espaciais, Avenida dos Astronautas 1758, S\~ao Jos\'e dos Campos, 12227-010, SP, Brazil\\
$^{5}$ \quad Astroinformatics, Heidelberg Institute for Theoretical Studies, Schloss-Wolfsbrunnenweg 35, 69118 Heidelberg, Germany}

\corres{Correspondence: spanda@lna.br}

\firstnote{CNPq Fellow} 

\abstract{The investigation of emission line regions within active galaxies (AGNs) has a rich and extensive history, now extending to the use of AGNs and quasars as ``standardizable'' cosmological indicators, shedding light on the evolution of our universe. As we enter the era of advanced observatories, such as the successful launch of JWST and the forthcoming Vera C. Rubin Observatory's Legacy Survey of Space and Time (LSST), the landscape of AGN exploration across cosmic epochs is poised for exciting advancements. In this work, we delve into recent developments in AGN variability research, anticipating the substantial influx of data facilitated by LSST. The article highlights recent strides made by the AGN Polish Consortium in their contributions to LSST. The piece emphasizes the role of quasars in cosmology, dissecting the intricacies of their calibration as standard candles. The primary focus centers on the relationship between the broad-line region size and luminosity, showcasing recent breakthroughs that enhance our comprehension of this correlation. These breakthroughs encompass a range of perspectives, including spectroscopic analyses, photoionization modeling, and collaborative investigations with other cosmological tools. The study further touches on select studies, underlining how the synergy of theoretical insights and advancements in observational capabilities has yielded deeper insights into these captivating cosmic entities.}


\begin{document}

\section{Introduction}

Active Galactic Nuclei (AGNs) are among the most luminous and powerful sources of energy in our Universe. AGNs are dynamic and powerful phenomena arising from the interaction between supermassive black holes and their surrounding environment \citep{Krolikbook}. The intense radiation originating from their very centers, how that interacts with the surrounding media, and their presence across a wide range of redshifts, make them crucial objects of study in understanding the cosmos at various scales and epochs \citep{
seyfert1943, schmidt63, greenstein63, greenstein_schmidt64, souffrin1969, lynden-bell1969, weedman1970, khachikian1971, weedman1976, weedman1977, blandford_mckee82, schmidt_green83, peterson88, peterson93, kaspi2000, sulentic2000, peterson2004, vester06, Tadhunter_2008NewAR..52..227T, bentz2009, Antonucci_2012A&AT...27..557A, bentz13, kormendy_ho2013, dupu2014, du2015, netzer2015, dupu2016a, padovani2017, grier17, du2019, Popovic_2020OAst...29....1P,2021ApJ...906...32Z, 2021ApJS..252...15M, 2022ApJ...925..121W, 2022MNRAS.510..687R, 2022MNRAS.513.1801L, 2022hxga.book....4B, 2022A&A...663L...7S, 2022ApJ...940...20G, 2023ApJ...943...67S, 2022ApJS..263...42W, 2023PASA...40...10O, 2023MNRAS.520.2781K, 2023A&A...676A..71M, 2023ApJ...951L...5Y, 2023ApJ...952...44B, 2023A&A...672A.137L, 2023MNRAS.521.2954S, 2023MNRAS.523.1399D, 2023ApJ...956...81S, 2023ARA&A..61..373F, 2023ApJ...957L...7K, 2023MNRAS.tmp.3133J, 2023arXiv230611099A, 2023arXiv230902516W, 2023arXiv231103459J, 2023arXiv231103590R}. They are characterized by the presence of a supermassive black hole (SMBH), accreting material from their surroundings, which leads to intense emission across a wide range of wavelengths. AGNs play a vital part in shaping the evolution of galaxies in the Universe. As material falls in due to the gravitational pull of the black hole (BH), it forms an accretion disk (AD) - a flattened structure of matter spiraling inwards as it loses angular momentum. The disk becomes significantly hot, of the order of $\sim 10^5$ K, and emits radiation peaking in the X-ray and UV wavelengths. The accretion disk emission gets intercepted by the gas clouds that, in turn, get ionized and emit the broad emission lines. This region is known as the broad line region (BLR) since the lines are characterized by the ``broadness'' of their line profiles. The broadening of the lines is due to a Keplerian distribution with velocities of the order of thousands to a few tens of thousand kilometers per second. Beyond the BLR lies the Narrow Line Region (NLR), another region of ionized gas emitting narrower emission lines. The NLR is characterized by lower ionization levels and extends farther out from the central BH \citep{netzer2015, padovani2017}. 

The distinguishing emission features in the BLR and NLR led to further categorization of the AGNs, especially through detailed studies of the nearby ``Seyfert'' galaxies \citep{seyfert1943}. Two distinct categories of Seyfert galaxies were eventually established based on the characteristics of their emission lines in the spectrum: those galaxies that exhibit permitted emission lines that are notably broader than their forbidden emission lines are classified as ``Type 1''. In contrast, galaxies displaying permitted and forbidden lines with similar widths are labeled as ``Type 2'' \citep{weedman1970, khachikian1971}. Over the subsequent years, the number of recognized Seyfert galaxies surged, and successive investigations unveiled the remarkable features of these celestial objects: (1) They possess remarkably compact and luminous nuclear regions, and (2) their spectrum contains a variable ultraviolet and optical continuum that complements their unusually strong and broad emission lines \citep[see][for an early review]{weedman1977}. AGNs are further classified into two well-known categories based on the strength of their radio emission: ``radio-loud'' (RL) and ``radio-quiet'' (RQ) AGN, a classification initially applied in the context of quasars (historically identified by sources with B-band magnitude $M_B < -23$) \citep[see, for example,][]{strittmatter1980,peacock1986,kellermann1989,miller1990, 1993ARA&A..31..473A, 1994AJ....108.1163K, 1995ApJ...438...62W, 1999AJ....118.1169X, 2000ApJ...543L.111L, 2000A&ARv..10...81V, 2002ApJ...565...78B,  2007ApJ...658..815S, 2010ApJ...711...50T, 2016ApJ...831..168K, 2017NatAs...1E.194P, 2019NatAs...3..387P}. This categorization was based on two primary criteria: (1) radio flux density or luminosity \citep[as demonstrated in][]{peacock1986}, and (2) the ratio of radio-to-optical flux density or luminosity \citep[as illustrated in][]{schmidt1970}. A critical parameter in this classification is the ``radio loudness'', characterized by the radio-to-optical flux density ratio, commonly referred to as the Kellerman index \citep[as detailed in][]{kellermann1989}. This index facilitated the grouping of radio-detected sources into these traditional categories, a practice that continues in the literature, albeit primarily applicable to Type-1 AGN \citep{padovani2011, bonzini2013}. However, recent studies by \citet{padovani2016} and \citet{padovani2017} have emphasized the need for a more precise distinction. They propose that the two classes represent fundamentally distinct entities, with RL AGN emitting a substantial portion of their energy non-thermally, often in conjunction with powerful relativistic jets. In contrast, the multi-wavelength emission of RQ AGN is primarily dominated by thermal emission linked to the accretion disk. As a result, the primary physical discrepancy between these two categories is the presence or absence of strong relativistic jets. Hence, these sources should be viewed more appropriately in terms of ``jetted'' and ``non-jetted'' rather than adhering to the traditional RL-RQ dichotomy \citep{padovani2017}. In this review, we will focus on the ``non-jetted'' sources and their salient properties.

The 1990s marked the rise of comprehensive surveys, such as the Bright Quasar Survey \citep{schmidt_green83}, enabling detailed statistical analyses of these sources \citep[e.g.][]{borosongreen1992}. This opened up the way to group Type-1 AGNs with the estimation of the width of their \hb{} emission line profile. This gave rise to a new dichotomy in these sources - Population A (Pop. A) and Population B (Pop. B). The notion of these two quasar populations was introduced as an analogy to the simplified classification of stars \citep[e.g.][]{sulentic2000}. Stars were cataloged into seven principal spectral types (OBAFGKM) in the HR Diagram. Population A sources included local Narrow-Line Seyfert 1 galaxies (NLS1s) as well as more massive, highly accreting objects that are predominantly classified as radio-quiet or non-jetted sources \citep[e.g.][]{marzianisulentic2014, padovani2017, Marziani_Panda_2023Galaxies}. These sources demonstrate FWHM(\hb{}) estimates of $\leq$ 4000 \kms{}. Notably, the \hb{} profile shape for Population A sources is typically Lorentzian-like \citep[e.g.][]{sul02,zamfiretal10,mar18}. On the other hand, Population B sources are noted to exhibit broader FWHM(\hb{}) values ($\geq$ 4000 \kms{}), and are frequently found to be ``jetted'' objects \citep[e.g.][]{padovani2017, Marziani_2021Univ}. Their \hb{} profiles are better described by Gaussian functions, and for those with even higher FWHMs, double Gaussian profiles characteristic of disk-like structures are observed in Balmer lines \citep[see e.g.,][]{2023ApJ...953L...3D}. Addressing the significance of the strict FWHM(\hb{}) threshold, studies by \citep[e.g.][]{sulentic2000,mar18, Panda_etal_2019QMS} indicate that AGN properties (such as the outflow intensity, accretion rate and metal content) exhibit notable shifts around this broader linewidth cutoff (at around 4000 \kms{}), mirroring the earlier division seen in Seyfert galaxies.

In this contribution, we group the studies based on the sub-structures that they are concerned with, i.e., Section \ref{sec:2} deals with our efforts to understand and utilize the information from the AD and the BLR, while Section \ref{sec:3} primarily highlights our efforts to probe the NLR using high transmission spectroscopy in the optical and near-infrared (NIR) regimes. AGNs offer insights into the Universe's structure, evolution, and fundamental properties. They can serve as distance indicators, probe large-scale structures, and contribute to our understanding of dark matter, dark energy, and galaxy formation. We discuss how AGNs that are and will be probed at high redshifts bridge the studies of the local and early Universe. We discuss some of these aspects in Section \ref{sec:4}. The review concentrates on results from recently published works led by me and my collaborators. Hence, we don't intend to provide an exhaustive coverage of the literature in this review. We highlight the essential references throughout the text in this work to guide the readers for further reading.

We highlight some recent progress in the studies of accretion disk structure and the development of a database of AGN SEDs, especially for the sources accreting at or above the Eddington limit (see Section \ref{sec:2}). We discuss the effectiveness of the BLR radius vs. AGN luminosity relation(s) in providing us with an independent way to estimate the luminosity distances to these cosmic sources, help build the Hubble diagram, and eventually test and retrieve cosmological parameters of our Universe (see Section \ref{sec:4.1}). Next, we propose a novel BH mass scaling relation traced by the intensities of high-ionization, forbidden emission lines (known as coronal lines) that we have discovered (see Section \ref{sec:3}). These coronal lines, a bulk of them produced in the NLR, are shown to evolve with the change in the underlying accretion disk structure, which is the primary source of the photons that produce emission of such lines (see Section \ref{sec:3.2}). We discuss our new-found BH mass scaling relation and how they can benefit from the up-and-coming JWST spectroscopic datasets.  Wherever necessary, we provide notes for future studies in these aforementioned sections. We then showcase our efforts in preparation for the upcoming Vera Rubin Observatory's Legacy Survey for Space and Time (LSST) in the context of the accretion disk and BLR time delay measurements using photometric monitoring and scrutinizing the survey strategies that are under discussion (see Sections \ref{sec:2.2} and \ref{sec:2.3}). We conclude with a summary of our work presented in this review with some notes for the future.

\section{Understanding accretion disk and BLR structure and emission}
\label{sec:2}

\label{sec:2.1}

\begin{sidewaysfigure}
    \centering
    \includegraphics[width=0.875\textwidth, height=0.45\textwidth]{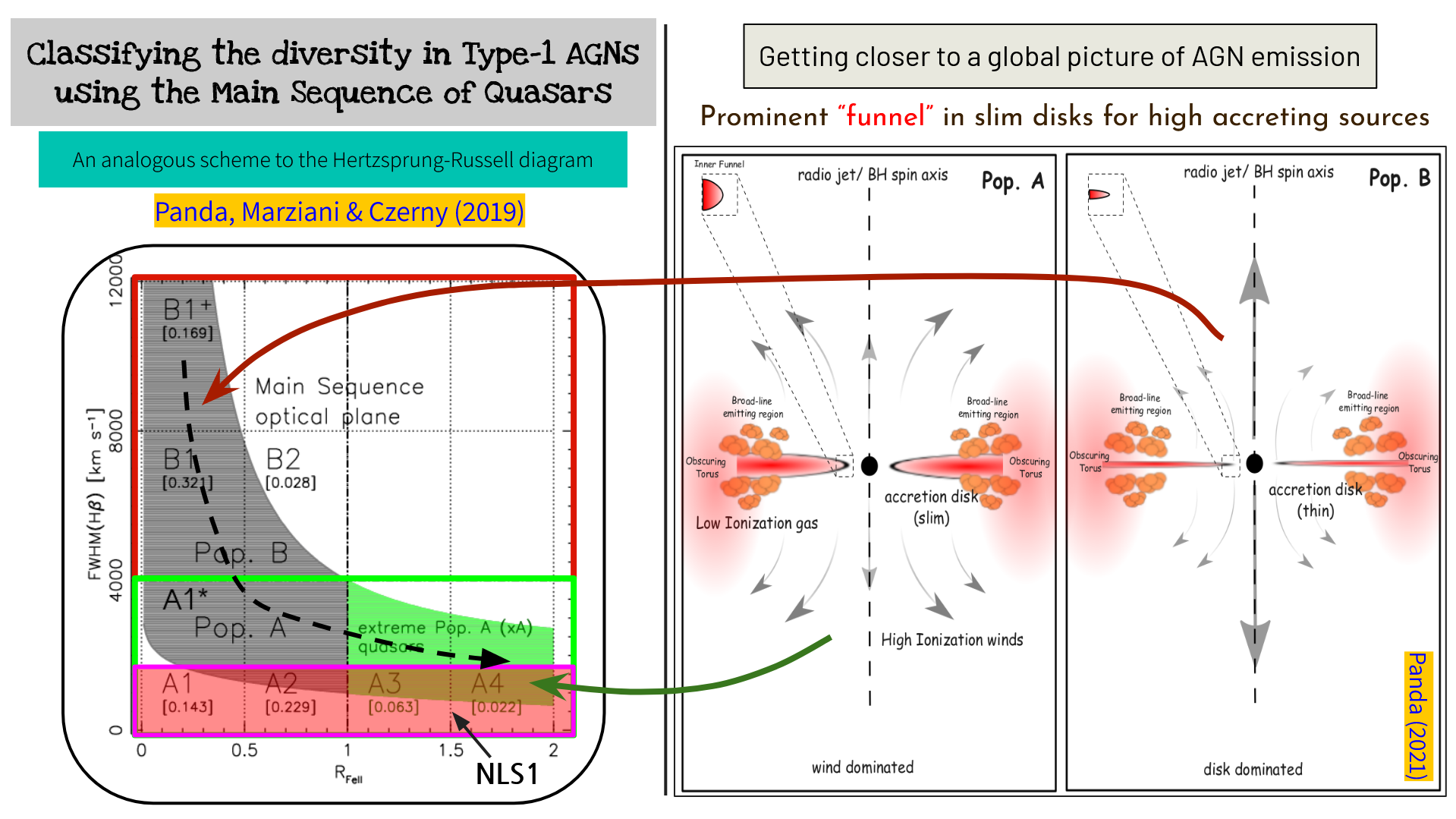}
    \caption{Top panel: Schematic view of the main sequence of quasars in the optical regime. The figure demonstrates the spectral types (e.g., A1, A2, B1, B2, B1+) based on the full-width at half maximum (FWHM) of the broad component of \hb{} versus the strength of the \feii{} flux in the optical (4434-4684\AA) to the \hb{} flux. Objects with FWHM(\hb{}) $\lesssim$ 2000 \kms{} are highlighted as Narrow-line Seyfert 1 (NLS1) galaxies, while those with FWHM(\hb{}) $\lesssim$ 4000 \kms{} belong to the Population A (Pop. A) and their counterparts with FWHM(\hb{}) $\gtrsim$ 4000 \kms{} are marked as Population B (Pop. B). Abridged figure based on \citet{Panda_etal_2019QMS}. Reproduced by permission of the AAS. Bottom panel: Schematic view of important components in the vicinity of supermassive black holes (SMBH) in AGNs. This illustration highlights the distinguishing characteristics of two different AGN types: Pop. A (left) and Pop. B (right). In both panels, an SMBH is located at the center, with a radio jet aligned along the BH's spin axis. Notably, the two panels contrast in terms of the accretion disk (AD) structure: the left panel showcases a slim AD, while the right panel features a thin disk conforming to the standard disk model (\citet{ss73}). Within this depiction, the BLR is divided into two regions: a high-ionization emitting region and a low-ionization emitting region. Additionally, the illustration includes the presence of a dusty, obscuring torus, with a covering factor nearly matching that of the low-ionization part of the BLR (\citet{Gaskell_2009}). It's worth noting the modifications introduced by the distinct accretion disk structures in the inset panels (see the top left corner of each panel). Specifically, the funnel-like shape within the innermost region of the slim accretion disk is notably more prominent in comparison to the scenario presented by the thin disk case. Abridged figure based on \citet{Panda_2021PhDT}.}
    \label{fig:qms_disks}
\end{sidewaysfigure}

In their comprehensive review, \citet{sul15} raised a pivotal question: \textit``{Do Population A and B represent two distinct quasar populations, or are they merely extremes on the same continuum? Could they be interconnected through a gradual transition in accretion mode?''} Figure \ref{fig:qms_disks} offers insights into this inquiry, suggesting that the latter scenario is operative. The figure underscores the transition in accretion disk structure from Population A to Population B, or vice versa, further elucidated in subsequent sections\footnote{The left panel of Figure \ref{fig:qms_disks} corresponds to the optical domain of Eigenvector 1, often referred to as the quasar main sequence. It defines the parameter space based on FWHM(\hb{}) and the strength of optical \feii{} emission (denoted as \rfe{}). The latter is determined by the ratio of \feii{} flux within 4434-4684 \AA{} (characterizing the blueward bump of the \hb{} profile) to the broad \hb{} flux.}.

The reverberation mapping (RM) technique uses the notion of the extra time taken by the continuum photon - one that gets intercepted by the broad line region clouds first before reaching us, and another that directly comes to us \citep{blandford_mckee82, peterson2004, kaspi_etal2000, bentz13, 1997ASSL..218...85N, 1999ApJ...526..579W, 2004PASP..116..465H, 2009NewAR..53..140G, 2009ApJ...702.1353D, 2010A&A...509A.106S, 2010ApJ...716..269W, 2010ApJ...721..715D, 2012ApJ...755...60G, 2014ApJ...793..108W, 2014MNRAS.445.3073P, 2015ApJS..217...26B, 2016ApJ...819..122Z, 2017ApJ...837..131P, 2017ApJ...840...97F, 2018ApJ...856....6D, 2018ApJ...866...75W, 2022MNRAS.509.4008P, 2023MNRAS.520.2009M, 2023MNRAS.522.4132Y}. This extra time is then translated to how far the BLR clouds are located from the continuum emitting source, i.e., \rblr{}. An important outcome of the investigations into RM is the establishment of an empirical power-law relationship between the \rblr{} and the AGN's continuum luminosity, often expressed as \rblr{} $\approx c \tau\ \propto $ \lopt{} $^\alpha$. Here, $\tau$\ represents the delay in time taken to respond to the variation in continuum for a given emission line, for example, \hb{}. The recent developments in RM have led to the expansion of the empirical \rblr{}-\lopt{} observational dataset \citep{kaspi2000, bentz13}, with a substantial increase in the total count, now exceeding 200 sources \citep{Shen_SDSS_2023arXiv230501014S}. This growth has been primarily driven by the inclusion of sources monitored under the SEAMBH project \citep{dupu2014, Wang2014_seambh, hu15, du2015, dupu2016a, dupu2018}, as well as contributions from the SDSS-RM campaigns \citep{grier17, Shen_2019ApJS..241...34S, Shen_SDSS_2023arXiv230501014S} and OzDES survey \citep[see e.g.,][]{Malik_OzDES_2023MNRAS.520.2009M}. However, this expansion has introduced a new challenge: an inherent dispersion in the \rblr{}-\lopt{} relation that has become apparent after incorporating these new sources (as depicted in Figure 1 of \citet{Panda_Marziani_2023}). Interestingly, some of these new sources have contributed to an increase in the dispersion for the relation, which correlates with the Eddington ratio. Studies have found that the sources that demonstrate larger deviation from the empirical \rblr{}-\lopt{} relation are found to have higher Eddington ratio \citep{martinez-aldama2019, du2019, Panda_Marziani_2023}.

\begin{sidewaysfigure}
    \centering
    \includegraphics[width=0.9\textwidth]{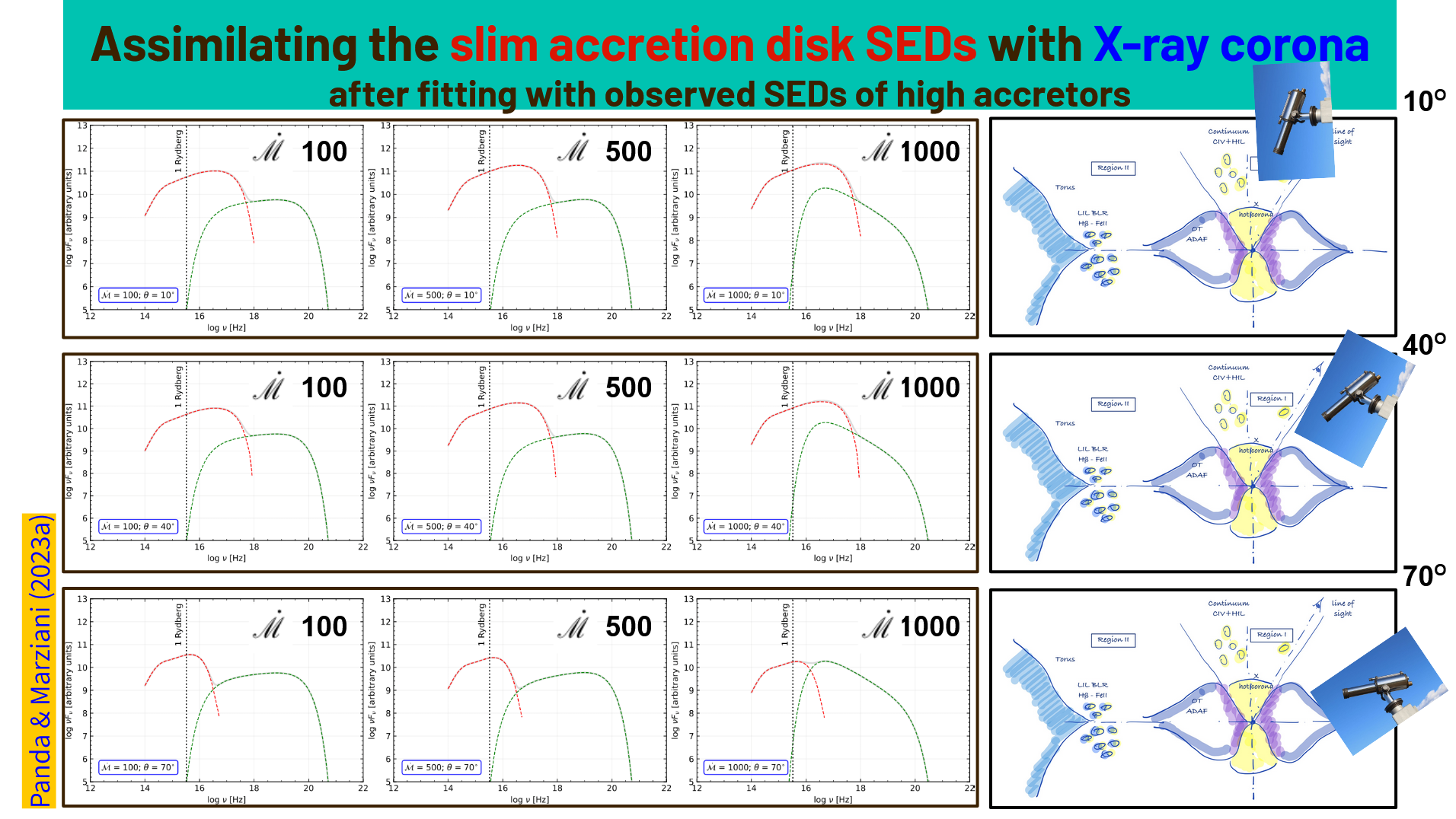}
    \caption{A collection of composite spectral energy distributions (SEDs) that combines a slim accretion disk (in red) and a high-energy, hot corona (in green). (From left to right:) The SEDs are shown as a function of increasing \mdot{} (or dimensionless accretion rate): (left panels) 100, (middle panels) 500, and (right panels) 1000. (From top to bottom:) Each \mdot{} case is accompanied by three distinct viewing angles (relative to an observer at a distance), denoted as \textit{i}: (upper panels) 10$^{\circ}$, (middle panels) 40$^{\circ}$, and (bottom panels) 70$^{\circ}$. The SEDs shown here assume a BH mass = 10$^8$ solar masses. The hot coronal emission is scaled to the slim disk. This adjustment employs the spectral index $\alpha_{\rm ox}$ that is obtained by comparing the modeled SEDs at the lowest inclination angle (i.e., 10$^{\circ}$) to the corresponding SEDs made directly from multi-wavelength observations. These observations are systematically arranged based on their \mdot{} values, as depicted in Figure 1 of \citet{ferland2020}. On the far right, a schematic view is provided, offering a conceptual representation of the inner region of a highly accreting SMBH. This representation is inspired by \citet{wang14}, although it is presented in a condensed format and not to scale. Abridged figure based on \citet{Panda_Marziani_SAB_2023}.}
    \label{fig:sed-models_slimdisks}
\end{sidewaysfigure}

\citet{martinez-aldama2019} investigated the connection of the increased scatter in the standard \rblr{}-\lopt{} relation with the Eddington ratio (\lbol{}/\ledd{}) for the first time. However, the Eddington ratio is a parameter estimated from the luminosity itself and thus creates circular reasoning. To alleviate this circularity, \citet{du2019} found that a separate observable parameter - the \feii{} emission strength in the optical regime, denoted as \rfe{}, can be used to correct for the added dispersion and the relation could be adjusted back with a slope close to 0.5. The introduction of the \rfe{} term is supported by earlier studies which have shown its close connection to the \lbol{}/\ledd{} \citep{sulentic2000, sh14, mar18, panda19b, du2019, martinez-aldama_2021}. Eventually, the introduction of the \rfe{} in the \rblr{}-\lopt{} relation compensates for the shorter time lags where sources with smaller \rblr{} sizes are the ones with strong \feii{} emission.

Addressing the shorter time lags observed in AGNs necessitates advancements in our photoionization modeling setup. This involves accounting for the Eddington ratio's influence on the continuum emission from the accretion disk and the resultant alterations in the net vertical distribution of the disk, particularly in locations near the innermost stable circular orbit (ISCO) of an SMBH. \citet{wang14} developed steady-state, analytical solutions for ``slim'' accretion disks, encompassing a broad range in accretion rates, from sub-Eddington to super-Eddington. In their models, they note the emergence of a funnel-like structure at the onset of the accretion disk, close to the SMBH. They attribute this to high accretion rates obtained from solutions of slim accretion disks. These solutions deviate \citep[][]{abramowicz88, sadowski2011, wang14} from the standard accretion disk scenario of \citet{ss73}. We present a schematic view in the right panel of Figure \ref{fig:qms_disks} which depicts the scenario. The highlighted structural changes to the disk not only impact the net anisotropic emission but also highlight the inclination effects due to the axisymmetric nature of these systems. Consequently, at high accretion rates, it is justified to account for viewing-angle-dependent anisotropy while modeling the continuum emission from the accretion disks. This scenario can explain well the dichotomy in the location and subsequent emission lines produced from the low- and high-ionization regions. For example, as shown in Figure \ref{fig:qms_disks}, the gas clouds emitting low-ionization lines are situated close to the disk plane, and effectively see a diminished disk luminosity relative to an observer with a line of sight nearly along the axis of rotation of the SMBH. This lowering in the flux incident on the BLR clouds in this region allows the clouds to encroach closer to the SMBH. This reduction in the location of the emitting regions aligns with observed estimates from RM campaigns \citep[for additional details, see]{panda_cafe2}.

In our modeling approach \citep{Panda_Marziani_SAB_2023, Panda_Marziani_2023}, we construct a database of composite SEDs comprising two main elements: a slim accretion disk (in red) and a hot corona (in green) in each panel of Figure \ref{fig:sed-models_slimdisks}). The two components are adjusted through the spectral index ($\alpha_{\rm ox}$), which represents the ratio of the flux at 2 keV to the flux at 2500\AA\ \citep{lusso2017}. The value for the spectral index is borrowed from the observed SEDs from \citet{ferland2020} where we match our model's composite SEDs viewed at 10$^{\circ}$ with these observed SEDs. We identify the best-fitting composite modeled SED for the corresponding cases of observed SED as a function of the dimensionless accretion rate (or \mdot{}, \citep{Wang2014_seambh, Panda_2022FrASS...950409P}). Figure \ref{fig:sed-models_slimdisks} (upper panels) demonstrates the best-fit modeled SEDs with increasing \mdot{}. We also explore the relation of the anisotropic emission on the viewing angle in these models. While we assume the hot corona originates from an isotropic radiation source and thus remains unaffected by changes in the viewing angle, the disk emission does vary with the viewing angle. This is highlighted in the right panels of Figure \ref{fig:sed-models_slimdisks}. The funnel-like structure gives a preferred direction for continuum radiation from the disk, more importantly for the photons that have origins near to the SMBH.

With the introduction of the anisotropic emission from a slim disk, we connect the BLR and the accretion structure where the high-ionization lines could be emitted within the funnel or, at the very least, in a location where the ionized gas receives the bulk of the continuum. Conversely, low-ionization lines might be exposed to a weaker continuum, necessitating the ionized clouds to move in to account for the diminished flux of the photons. A comprehensive assessment of outcomes from our photoionization modeling will be presented in an upcoming work.

\section{AGNs for Cosmology}
\label{sec:4}

AGNs play a significant role in cosmology by serving as valuable tools for various aspects of studying the universe's large-scale structure, evolution, and fundamental properties. AGNs with known luminosities, such as quasars, can serve as distance indicators in the universe. Since their intrinsic brightness is relatively consistent, their observed brightness can be used to estimate distances, aiding in measuring cosmological distances and understanding the expansion of the universe. In the next section, we discuss some salient details of this technique and highlight its shortcomings and how we as a community have devised ways to overcome these challenges to reinstate the use of quasars as standardizable distance indicators.

\subsection{Effectiveness of the Radius-Luminosity Relation(s)}
\label{sec:4.1}

\begin{sidewaysfigure}
    \centering
    \includegraphics[width=0.8\textwidth]{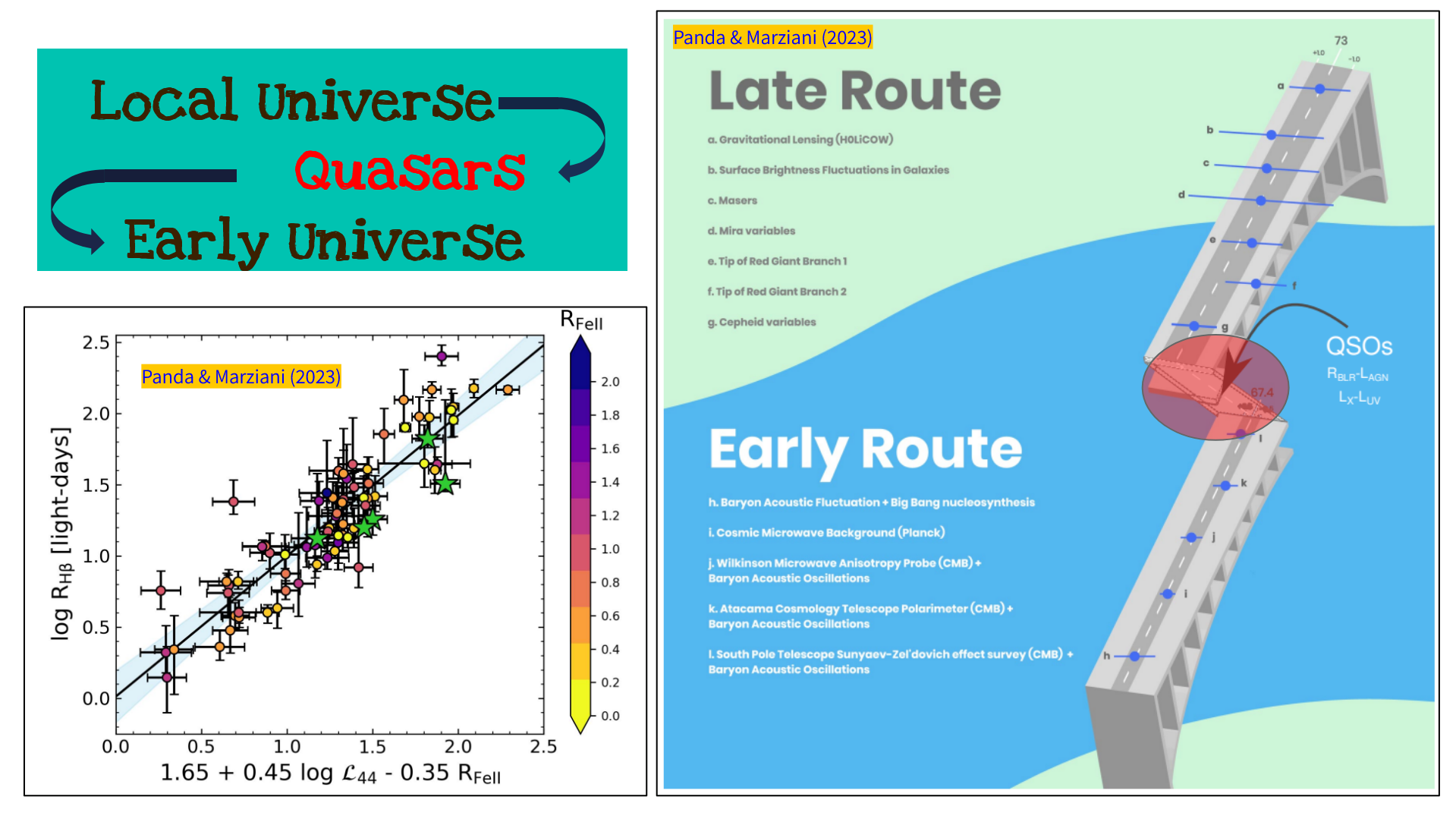}
    \caption{Collage demonstrating the use of quasars as standardizable distance indicators (Left:) A summarized rendition of Figure 5 from \citet{du2019} displays the updated relationship between \rblr{} versus \lopt{} and incorporating the parameter \rfe{} for 75 sources. The black solid line represents the best-fit relation, characterized by a $\rho$=0.894 and a p-value=3.466$\times 10^{-27}$. The region marked in blue shows the 99\% confidence interval around the best-fitting line, encompassing a scatter of approximately 0.196 dex. In this context, $\mathcal{L}_{44}$ denotes the \lopt{} normalized by 10$^{44}$ erg s$^{-1}$. Sources exceeding \lledd{} $>$ 3 are highlighted using green star symbols. Abridged figure based on \citet{Panda_Marziani_2023}. (Right:) This diagram illustrates the diverse range of methodologies utilized for determining the Hubble constant (${H_0}$), one that characterizes the expansion rate of the Universe. The observations can be divided into two categories: one involving investigations of the very early Universe (depicted in the lower half of the diagram), and the other focusing on the Universe's expansion estimated from the local distance indicators (shown in the upper half). On the right-hand side of the graphic, the letters associated with each technique (enumerated on the left) are placed along the bridge. Collating data from the various methods within the low-z Universe exploration gives an average ${H_0}$ estimate of 73 km s$^{-1}$ Mpc$^{-1}$. For the early Universe techniques, the combined ${H_0}$ value is determined to be 67.4 km s$^{-1}$ Mpc$^{-1}$. Selected quasar-based techniques, the radius-luminosity relation, and another - X-ray to UV luminosity relation are highlighted in the red circle. Abridged figure based on \citet{Panda_Marziani_2023}.}
    \label{fig:quasars_cosmo}
\end{sidewaysfigure}

An important outcome of the investigations into reverberation mapping is the establishment of an empirical relationship between the radius of the BLR measured from the time delay (\rblr{}) and the AGN's luminosity, often expressed as \rblr{} $\approx c \tau\ \propto $ \lopt{}$^\alpha$. Here, $\tau$\ represents the delay in light travel time responding to the variation in the continuum of a given emission line, typically \hb\footnote{This relationship supposes the radius of the BLR for H$\beta$ and the adjacent continuum luminosity at 5100\AA}. Notably, \citet{bentz13} conducted a study on 41 AGNs covering a luminosity range spanning four orders of magnitude and determined a best-fit power-law slope value of $\alpha = 0.533^{+0.035}_{-0.033}$. This result is remarkably similar to $\alpha = 0.5$ (from theoretical expectations), which is necessary to preserve spectral similarity \citep{wills1985, davidson_1977, osterbrock_ferland06}. By combining this \rblr{}-\lopt{} relation with the line profile widths of broad emission lines estimated from spectroscopy made at single or multiple epochs, one can gauge \mbh{}, rendering the relation particularly valuable for extensive surveys across wide redshift ranges \citep{vester06, s11}.

Dedicated monitoring has significantly expanded the population in the \rblr{}-\lopt{} observational domain, surpassing a count of 100 (see \citep{martinez-aldama2019, panda_2019_frontiers, Shen_SDSS_2023arXiv230501014S} for more details). Notably, this was achieved through contributions from projects such as the SEAMBH project (Super-Eddington Accreting Massive Black Holes; \citep{dupu2014, Wang2014_seambh, hu15, du2015, dupu2016a, dupu2018}), the SDSS-RM campaigns \citep{grier17, Shen_2019ApJS..241...34S}, and individual photometric reverberation campaigns \citep{Haas2011, Pozonunez12, 2023ApJ...949...22M}. However, these developments have introduced a new issue, i.e., the increased scatter in the \rblr{}-\lopt{} relation arising after the inclusion of these new sources. The left panel of Figure 2 in \citet{Panda_Marziani_2023}, which is a modified version from \citet{martinez-aldama2019} and \citet{panda_2019_frontiers}, illustrates the \rblr{}-\lopt{} observational space for 117 AGNs that underwent reverberation mapping. Color-coded according to their Eddington ratios (\lbol{}/\ledd{}), this plot reveals the best-fit relation for the aforementioned dataset as log R$_{\rm H\beta}$ = 0.387$\times$($\log\;{\rm L_{5100}}$) - 15.702. It exhibits a Spearman's correlation coefficient ($\rho$) of 0.733 and a p-value of $2.733 \times 10^{-21}$, leading to a relation with a significantly shallower slope than the previous findings of \citet{bentz13}. This prompts questioning the validity of the empirical \rblr{}-\lopt{} relation.

Interestingly, the sources that contribute to the increased dispersion in the relation exhibit a discernible relation with the Eddington ratio. Namely, sources deviating more from the empirical \rblr{}-\lopt{} relation tend to possess higher Eddington ratios. In \citet{martinez-aldama2019}, it was demonstrated that this dispersion can be reconciled within the standard \rblr{}-\lopt{} relation by introducing a dependency on the Eddington ratio (\lbol{}/\ledd{}). This phenomenon is highlighted in the right panel of Figure 2 in \citet{Panda_Marziani_2023}, where the offset ($\Delta$\rblr{}) between the observed time delays and those calculated using the empirical \rblr{}-\lopt{} relation \citep[$\Delta$\rblr{},][]{bentz13}, is plotted against their respective \lopt{} values for the sources presented in the left panel. Notably, there is a distinct reduction in $\Delta$\rblr{} values for higher Eddington ratio sources, suggesting that objects with smaller observed time lags, particularly at the high-luminosity end of the relation, are accreting at or about the Eddington limit. Consequently, these objects are anticipated to host SMBHs with lower \mbh{} \citep{dupu2014, dupu2016a}. While certain high-Eddington sources exhibit reduced \rblr{} sizes, some sources with comparable \rblr{} sizes to those predicted by \citet{bentz13} remain despite their high Eddington nature. Building upon this, \citet{du2019} further demonstrated that the relation can be restored to its original slope of $\sim$0.5 by introducing an additional correction term. This supplementary term is based on the optical \feii{} strength measured to the H$\beta$ emission (\rfe{}), that has been validated as a reliable observational proxy for the Eddington ratio in previous studies \citep{sulentic2000, sh14, mar18, panda19b, du2019, martinez-aldama_2021}, as previously discussed. The analytical relation thus becomes \citep{du2019}: 

\begin{equation}
\log\left(\frac{{\rm R_{BLR}}}{1\;\mathrm{light-day}}\right) = \kappa + \alpha{}\log\left(\frac{{\rm L_{5100}}}{10^{44}\, \mathrm{erg\ s^{-1}}}\right) + \gamma{}{\rm R_{FeII}}
\label{eq:rcorr}
\end{equation}

By incorporating this correction in terms of a measurable quantity, \rfe{}, we circumvent the circularity issue that arose when the correction was explicitly based on the Eddington ratio \citep{martinez-aldama2019}. This approach provides an independent estimation of the luminosity distance (D$_{\rm L}$). Consequently, we can build a Hubble diagram using quasars for which we know both their luminosity distances and redshifts. This setup allows us to test the validity of the standard cosmological model as well as alternative models \citep[see e.g.][]{Haas2011, watson2011, czerny2013}. Notably, super-Eddington accreting sources are anticipated to be favorably selected as redshift increases, particularly in a flux-limited sample \citep{sulenticetal14}. This selection bias, combined with intrinsic evolution in the Eddington ratio \citep[e.g.,][]{cavalierevittorini00, hopkinsetal06}, often results in high-redshift quasar spectra resembling extreme Population A spectra found at lower redshifts \citep{sulentic2000}. Thus, the inclusion of such sources is pivotal to extend the \rblr{}-\lopt{} relation to the higher luminosity regime, as illustrated in the left panel of Figure \ref{fig:quasars_cosmo}, where the parameters $\kappa$=1.65$\pm$0.06, $\alpha$=0.45$\pm$0.03, and $\gamma$=-0.35$\pm$0.08 are employed.

The incorporation of \rfe{}, particularly for objects exhibiting intense \feii{} emission, effectively compensates for the shorter time delays, thereby accounting for the smaller \rblr{} sizes observed. Furthermore, this approach enables the recovery of a slope ($\alpha$) that aligns more closely with the theoretical predictions, specifically 0.5 \citep{davidson_1977, Davidson_Netzer_1979}. Consequently, we can confidently employ the modified \rblr{}-\lopt{} relation (inclusive of the correction term involving \rfe{}) to determine the source's monochromatic luminosity, independently. By combining the knowledge from directly observed flux with this approach, we can eventually derive accurate estimates for the luminosity distance to the target source and avoid the circularity problem: 

\begin{equation}
{\rm D_L} = \sqrt{{\rm L_{5100}}/4\pi {\rm F}} \propto \frac{c\tau}{\sqrt{4\pi F}}
\label{eq:dl}
\end{equation}

\section{Coronal lines as BH mass tracers using optical and NIR spectroscopy}
\label{sec:3}

The Narrow-Line Region (NLR) is another important component of AGNs, distinct from the BLR and the accretion disk. It is located farther out from the black hole than the BLR and is characterized by narrower emission lines in its spectrum indicating that the gas in the NLR moves at lower velocities in case of a virial velocity field. Due to the characteristic low particle densities ($\lesssim 10^7$ \cc{}), the NLR also emits forbidden lines (e.g., the coronal lines) of ionized elements like oxygen, nitrogen, and sulfur. The emission lines from the NLR provide valuable diagnostic tools for studying the physical conditions of the gas, e.g., its density, temperature, and metallicity. The ratios of different emission lines can be used to deduce the properties of the ionizing radiation field and the gas dynamics.

\begin{figure}[!htb]
    \centering
    \includegraphics[width=0.9\textwidth]{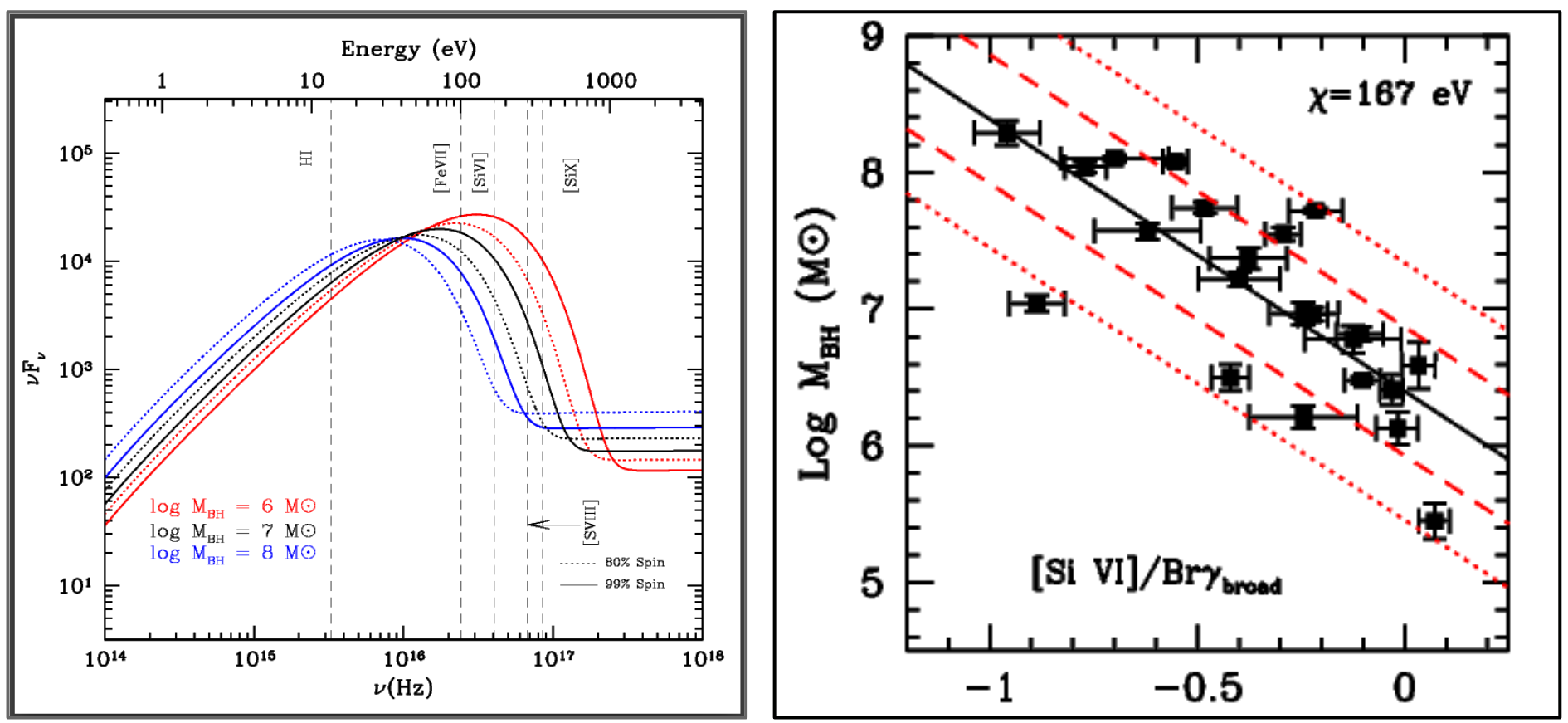}\\
    \includegraphics[width=0.9\textwidth]{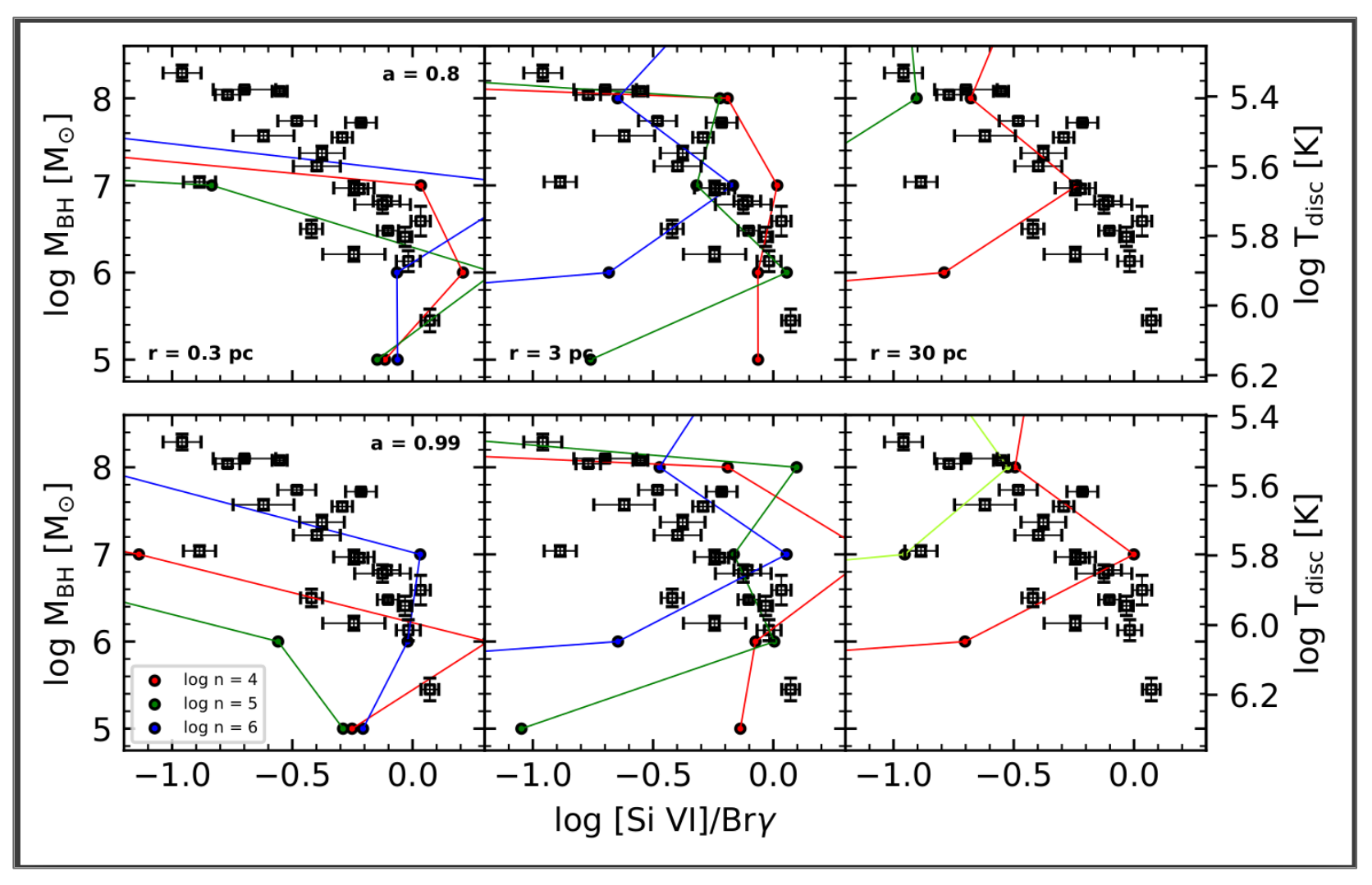}
    \caption{Coronal line (CL) emission driven by SMBH-dependent properties in an AGN SED (Top left:) In this study, the generic ionizing continuum for AGNs is formulated according to Eq. \ref{eq:sed}, accommodating the range of BH masses taken from the AGN sample \citep{bentz+15} where the masses are derived from the \hb{}-based reverberation mapping (see top right panel). For each BH mass, an associated temperature of the accretion disk ($T_{\rm disk}$) is considered, adhering to the thin disk approximation as delineated in Eq. \ref{eq:tbbb}. This representation results in a series of curves, where each curve corresponds to a distinct BH mass. We also include the dependence of the BH spin in these SEDs, i.e., two values of the BH spin: a = 0.8 (dashed) and a = 0.99 are incorporated. The analysis encompasses three representative BH masses: 10$^6$ (in red), 10$^7$ (in black), and 10$^8$ (in blue). The BH masses are reported in solar masses. To provide context, vertical dashed lines are employed to signify the ionization potentials (IPs) of the lines incorporated in the analysis; (Top right:) Observed [\ion{Si}{vi}]~1.963$\mu$m/Br$\gamma_{\rm broad}$ ratio (denoted by R) versus BH mass for the sources analyzed in \citet{Prieto_etal_2022}. The linear best-fit is shown using the black line and the 1$\sigma$ and 2$\sigma$ deviation are shown using the red-dashed and -dotted lines, respectively; (Bottom:) CLOUDY predictions for [\ion{Si}{vi}]/Br$\gamma_{\rm broad}$ vs BH mass using the ionizing continua shown in the top-right panel. Models are simulated for local cloud densities, n$_{\rm H}$ = 10$^4$ \cc{} (red), 10$^5$ \cc{} (green), and 10$^6$ \cc{} (blue). Each sub-panel shows the results from the {\sc cloudy} modeling for varying distance from the ionizing source, (from left to right) \textit{r} = 0.3, 3, and 30 pc, respectively. We consider two BH spins: a = 0.8 (upper panel) and a = 0.99 (lower panel). We depict the corresponding $T_{\rm disk}$ per each BH mass, on the right side. Observed data are marked in black squares. Abridged figure based on \citet{Prieto_etal_2022}. Reproduced with permission from The Royal Astronomical Society.}
    \label{fig:coronal-lines}
\end{figure}

Coronal Lines (CLs) are a special class of emission lines with a majority of them originating in the NLR and are characterized by their elevated ionization potentials (IPs), spanning from approximately 50 eV up to a few hundred eV. This unique property renders them exceptional indicators of the ionizing continuum \citep{1992ApJ...387..522L, 2002MNRAS.331..154R, 2021ApJ...922..155M, 2022MNRAS.516.4397K, 2023ApJS..265...21R, 2023MNRAS.525.1568S}. While they may appear fainter compared to the conventional medium-ionization lines frequently employed for photoionization analysis, advancements in high angular resolution techniques for nearby AGNs have unveiled that CLs, notably those observable in the NIR spectrum, stand out as some of the most prominent features \citep[e.g.][]{marconi1994, muller-sanchez+11, rodriguez+17, gravity2020}. The selection of objects in this study is based on two main criteria: \mbh{} estimated through reverberation mapping and the availability of single-epoch optical and/or NIR spectra with high-quality measurements of the CLs. The initial basis confines that our sources belong to the Type-1 AGNs exclusively, while the latter criterion aims to mitigate potential issues stemming from variability. While a preference is given to sources offering both optical and NIR spectra, this specific requirement could not always be met. The utilized coronal lines include [\ion{Fe}{vii}]~$\lambda$6087~\AA\ in the optical spectrum and [S\,{\sc viii}]~0.991~$\mu$m, [Si\,{\sc x}]~1.432~$\mu$m, and [Si\,{\sc vi}]~1.964~$\mu$m in the NIR spectrum. These lines are recognized as among the most potent CLs in AGNs \citep{reunanen03, rodriguez+11} and span a substantial ionization potential (IP) range of 100 to 350 eV. Additionally, H\,{\sc ii} lines, specifically H$\beta$, Pa$\beta$, and Br$\gamma$, are incorporated. Collectively, this set of lines captures the ionizing continuum over a range spanning from 13.6 to 351 eV. The preference for NIR CLs is attributed to their reduced susceptibility to extinction, while the choice of optical CL [\ion{Fe}{vii}] is attributed to its robust strength, modest extinction, and an IP in the vicinity of [Si\,{\sc vi}]. The final selection comprises 31 AGNs (Table 1 in \citet{Prieto_etal_2022}). This dataset encompasses BH masses, primarily sourced from the compilation by \citet{bentz+15}, along with CL ratios extracted from spectra studied in \citet{Prieto_etal_2022} or derived from previously documented spectroscopic data, along with the corresponding data sources. The majority of the line flux ratios in the near-infrared (NIR) utilized in this study are derived from \citet{riffel+06}. In cases where the mentioned publication did not provide data for specific targets, we acquired observations through the use of the Gemini Near-Infrared Spectrograph (GNIRS) on the Gemini North Observatory, or the ARCOiRIS/TripleSpec4 near-infrared spectrograph, which was installed on either the 4m Blanco or 4m SOAR telescopes.

We demonstrate the relationship of the \mbh{} with a prominent CL, [\ion{Si}{vi}]~1.963$\mu$m which has an IP [\ion{Si}{vi}] = 167 eV\footnote{This CL ranks as one of the most prevalent and luminous features observed in the spectra of AGNs, as discussed in previous studies \citep{rodriguez+11}. For a comprehensive summary of our findings, we direct readers to the extensive analysis provided by  \citet{Prieto_etal_2022}.} after normalizing it to the nearest H\,{\sc i} broad line emission, i.e., Br$\gamma$, where the dependence to the \mbh{} was found to be the strongest. We observe a tight correlation between \mbh{} and the CL ratio [\ion{Si}{vi}]/Br$\gamma_{\rm broad}$. 
Figure~\ref{fig:coronal-lines} introduces a novel diagnostic diagram, showcasing the correlation between the black hole (BH) mass of the objects in our sample and the ratio of [\ion{Si}{vi}]~1.963$\mu$m to Br$\gamma_{\rm broad}$\footnote{We note in passing that there is a good agreement between the FWHMs obtained for Br$\gamma_{\rm broad}$ and \hb{} where the Br$\gamma_{\rm broad}$-based FWHMs are narrower by $\sim$300 \kms{} \citep[see e.g.,][]{2008ApJS..174..282L}}. This ratio exhibits a distinct trend with BH mass spanning three orders of magnitude. A linear regression analysis applied to the data yields the following result:
\begin{equation}
    \log M_{\rm BH} = (6.40\pm 0.17) - (1.99\pm 0.37) \times \log \left(\frac{[{\rm Si\,{\sc VI}}]}{{\rm Br}\gamma_{\rm broad}}\right),
\end{equation}
which has a 1$\sigma$ scatter of 0.235 dex. We used the {\sc LtsFit} package\footnote{\href{http://www-astro.physics.ox.ac.uk/~mxc/software/\#lts}{http://www-astro.physics.ox.ac.uk/mxc/software/lts}} \citep{capellari+13} for the regression analysis accounting also for the errors in the two parameters. The Pearson correlation coefficient for the correlation is $r$ = -0.76, with a \textit{p}-value = 3.8$\times 10^{-5}$. In contrast, when examining 49 galactic bulges with directly measured dynamical black hole (BH) masses, a dispersion of 0.22 dex is revealed in the $M-\sigma$ relation \citep{gultekin+09}. It's worth noting that the intrinsic scatter in mass-luminosity relationships is approximately 40\%, as reported by  \citet{kaspi_etal2005}. This variability is primarily attributed to disparities in the shape of the optical-UV continuum.

The current scaling relation for BH masses is constrained to Type-1 AGNs, encompassing narrow-line Seyfert galaxies as well. This limitation arises from the necessity to include only reliably determined BH masses and to normalize them against broad \ion{H}{i} gas measurements. Nevertheless, efforts are underway to explore the feasibility of extending this relation to Type-2, obscured AGNs. The newly proposed scaling relation offers an efficient and theoretically-inspired alternative for estimating \mbh{} using single epoch spectroscopic data. This approach eliminates the need for extensive telescope time, as required in techniques like reverberation mapping, or the necessity for highly precise absolute flux calibration, which is essential in methods such as the continuum luminosity approach. By making use of the capabilities of the JWST and extensive infrared surveys, this method has the promise of efficiently determining the masses of large samples of AGNs. For a detailed examination of the observed spectra of the sources and comprehensive photoionization modeling to validate the underlying principles of the observed scaling relationship, we recommend referring to the work conducted by \citet{Prieto_etal_2022}.

\subsection{Standard disk predictions and radiative transfer modeling}
\label{sec:3.2}

The correlation observed between coronal line (CL) emission and BH mass prompted an investigation into the potential connection between the aforementioned CL emission and predictions derived from standard accretion disk theory. This exploration aimed to establish a link between CL emission and a key attribute of the accretion disk—specifically, the disk's temperature peak. To test this hypothesis, the photoionization code CLOUDY \citep[v17.02, ][]{ferland+17} was employed to generate a range of synthetic quasar spectra encompassing varied physical conditions. The primary objective was to assess whether the observed CL line ratios exhibit a discernible dependence on $T_{\rm{disk}}$, which is introduced to CLOUDY through the ionizing continua depicted in Figure \ref{fig:coronal-lines} (upper left panel). In this figure, the CLs are attuned to distinct energy ranges within the AGN ionizing continuum. This continuum arises from a composite of two components: a standard Shakura-Sunyaev (SS) accretion disk \citep{ss73}, describing the spectral transition from ultraviolet (UV) to soft X-rays involves capturing a power-law segment featuring both lower and upper energy limits, addressing the spectrum's rise at higher energies. It follows the equation

\begin{equation}
    F_{\nu} = \nu^{\alpha_{uv}}\exp{\left(\frac{-h\nu}{kT_{disk}}\right)}\exp{\left(\frac{-kT_{IR}}{h\nu}\right)} + a\nu^{\alpha_x}
\label{eq:sed}    
\end{equation}

The first component is an expression of the SS disk, formulated as an exponential function featuring a cutoff at the effective temperature of the disk, $T_{\rm{disk}}$, coupled with a power-law segment characterized by $\alpha_{\rm uv}$ = 0.33 to account for the disk's low-energy tail. The lower boundary of the disk's energy range is defined by an infrared-exponential component with a cutoff at 0.01 Ryd. In the high-energy regime, the distribution is described by a broken power law with a spectral index denoted as $\alpha_{\rm x}$ = -1, which exhibits a cutoff at 100 keV. The relationship between the SS disk and the high-energy power law is governed by the parameter represented as \textit{a} in Equation \ref{eq:sed}, indicating the ratio of luminosities at 2 keV and 2500 \AA. This ratio is expressed through a power-law function with a spectral index $\alpha_{ox}$ set at -1.4. Except for $\alpha_{\rm uv}$ = 0.33, all other parameters mentioned here adhere to the general AGN continuum utilized in CLOUDY \citep{mathewsferland87}. In the ionizing continuum presented in Figure \ref{fig:coronal-lines} (upper left panel), the peak emission effectively corresponds to $T_{\rm{disk}}$, with the disk's temperature increasing as the BH mass decreases, aligning with predictions from standard accretion disk theory. Importantly, it is worth noting that the ionization potentials of the coronal lines align closely with the peak temperatures, or slightly beyond, depending on the BH mass. If the accretion disk primarily acts as the source of photons for ionizing the coronal gas, we can anticipate a relationship between the strength of coronal lines and the temperature of the accretion disk, denoted as $T_{\rm{disk}}$. Given the relationship between $T_{\rm{disk}}$ and BH mass, this connection might consequently result in a correlation between the strength of coronal lines and BH mass. We follow the standard accretion disk prescription, and employing it for a BH in Kerr metric, the disk temperature, $T_{\rm disk}$, can be written as (\citet{frank+02,2016ASSL..440....1L} prescription is used):

\begin{equation}
\begin{split}
 T_{\rm disk} = & 3.4
 \times 10^5 K ~ \left(\frac{M_{\rm BH}}{10^8 M_{\odot}}\right)^{1/4} \times \left(\frac{\left(\frac{dM}{dt}\right)}{0.1}\right)^{1/4}\\
 & \times \left(\frac{\eta}{0.26}\right)^{-1/4} \times \left(\frac{R_{G,in}}{1.4}\right)^{-3/4}
 \label{eq:tbbb}
\end{split}
\end{equation}

Here, \mbh{} represents the black hole mass, $(dM/dt)$ signifies the accretion rate in Eddington units, $\eta$ stands for the black hole accretion efficiency, and $R_{G,in}$ represents the innermost stable circular orbit (ISCO), measured in units of gravitational radii, $R_G = GM_{\rm BH}/c^2$ units (where \textit{G} represents the gravitational constant and \textit{c} is the speed of light). This equation is normalized for a BH mass of $M_{\rm BH}$ = 10$^8$ M$_{\odot}$, accreting at 10\% of the Eddington limit (i.e., $dM/dt$ $\sim$ 0.1 $(dM/dt)_{Edd}$), and $\eta$ = 0.26, that corresponds to a BH that is maximally spinning, i.e., $a = 0.99 cJ/G$\mbh{}$^2$ (hereafter referred to as a = 0.99, where $J$\ is the spin angular momentum), where the rotation is in clockwise direction. Recent estimations of BH spin from various methods tend to converge towards values close to 1 \citep[see][for a comprehensive compilation]{reynolds2019}. As the spin of the object increases, the ISCO contracts, resulting in higher values for the accretion disk temperature, denoted as $T_{\rm disk}$. When the condition of co-rotation is met, the disk temperature reaches its maximum value.

The ionizing continua follow the correlation between the accretion disk temperature ($T_{\rm disk}$) and BH mass, as described in Equation \ref{eq:tbbb}. Curves are presented for three specific BH masses and two spin values, namely 0.8 and 0.99. Spin values below 0.8 result in minimal deviations in $T_{\sc disk}$ when compared to the a = 0.8 scenario, and are consequently omitted from presentation \citep[see, for instance, a comprehensive study on spin parameter space by][]{campitiello+19}.

With such an incident continuum as the predominant contributor to ionize the emitting regions, and considering an appropriate range of local cloud densities ($n_{\rm e} \sim 10^{4-6}$ cm$^{-3}$) as well as cloud distances (ranging from 0.3 pc to approximately 30 pc) to ensure the continuance of CLs, photoionization models conducted using {\sc CLOUDY} \citep{2017RMxAA..53..385F} yield a reasonably accurate depiction of the correlation between \mbh{} and the [Si {\sc vi}]/Br$\gamma_{\rm broad}$ (see bottom panel of Figure \ref{fig:coronal-lines}).

It is discernible that the peak emission from the disk progressively approaches the ionization potentials of $Si^{+5}$ and $Fe^{+6}$ as the BH mass decreases and the spin increases. However, as we transition to higher BH masses ($10^7 - 10^8$ \msun{}) or beyond, the peak energies obtained from the accretion disk alone is insufficient for the ions with higher IPs, $Si^{+9}$ and $S^{+7}$, with $T_{disk}$ becoming cooler as the BH mass rises. This is where we bring in the SEDs that include the warm comptonizing component reflected in the soft X-ray regime. We make a re-run of the CLOUDY models by replacing the incident continuum shape with these modified accretion disk SEDs. We notice a clear increase in the flux available at the locations for the higher ionization CLs, i.e., $Si^{+9}$ and $S^{+7}$ ions. These new models provide better agreement to the observed correlations between the CL ratios and the BH mass for these other ionic species. The results of the $Si^{+5}$ and $Fe^{+6}$ ions are relatively unaffected.

\section{Getting ready for LSST: predictions for BLR and AD time-lag recovery}
\subsection{BLR time-lag recovery expectations from LSST}
\label{sec:2.3}

The Vera C. Rubin Observatory's Legacy Survey of Space and Time \citep[LSST; ][]{LSST_Ivezic_etal_2019, LSST_Bianco_etal_2022, 2009arXiv0912.0201L, 2012arXiv1211.0310L, 2017arXiv170804058L, 2018ApJS..236....9N, 2021ApJS..253...31L} is poised to revolutionize our understanding of the Universe by delivering an unparalleled volume of data across various fields. Observations are set to commence relatively soon, likely around mid-2024, underscoring the significance of optimizing data utilization, especially during the initial phases of operation. Amidst its myriad outcomes, LSST is anticipated to unveil up to ten million quasars, paving the way for extensive reverberation monitoring of AGNs over a redshift range spanning from 0 to 7 \citep{shen_X_2020, LSST_Kovacevic2022}.

Broad emission lines are a hallmark of nearly all luminous AGN \citep[see e.g.][for a comprehensive review]{Krolikbook}. Nonetheless, the exact location and configuration of the corresponding region, known as the Broad Line Region (BLR), remain largely unresolved, except for recent infrared observations of three active galactic nuclei (AGN) using the GRAVITY Very Large Telescope Interferometer (VLTI) instrument. These AGN include 3C 273 (a well-known radio-loud source, as seen in \citealt{GRAVITY_3C273}), and two radio-quiet sources: NGC 3783 (as observed in \citealt{GRAVITYNGC3783}), and IRAS 09149-6206 (as reported in \citealt{GRAVITY_iras}). For most other objects, our understanding of BLR structure and dynamics heavily relies on reverberation monitoring. While the LSST will provide photometric light curves for numerous AGNs, the translation from photometry to line time delays is intricate and demands precision. It is essential to assess the reliability of measurements, taking into account source luminosity, redshift, and observing cadence, particularly when using LSST's photometric bands. As a response, we have developed a simulation tool aimed at optimizing measurement success across the entire survey, the first two years, and the inaugural year. This optimization effort encompasses the Main Survey (MS) and the planned Deep Drilling Fields (DDFs). The LSST's main survey (MS) will encompass roughly 18,000 square degrees of the southern sky, employing 6 filters. Each location within this survey area is anticipated to receive approximately 825 observations over the entire campaign duration. The expected magnitude limits, with a signal-to-noise ratio (SNR) greater than 5$\sigma$, are r $<$ 24.5 in individual images and r $<$ 27.8 when considering the fully stacked data. Apart from the Rubin Observatory's extensive main survey, conducted using a ``universal cadence'' a notable portion of Rubin's observing time, approximately 10\%, will be allocated to various additional programs. This allocation includes the intensive observation of specific DDFs, namely COSMOS, Extended Chandra Deep Field-South (ECDFS), XMM-LSS, ELAIS S1, and Euclid Deep Field Survey (EDFSa and EDFSb). These DDFs will receive more comprehensive coverage and more frequent temporal sampling, particularly in some of the LSST's ugrizy filters, in comparison to the standard observations of the broader sky \citep[see e.g.,][]{2018arXiv181106542B, LSST_Kovacevic2022}. The LSST DDFs will have the deepest coadded depths of the survey \citep{2018arXiv181106542B}, and are expected to exceed those achieved by 8-meter class observatories in the same areas \citep[see][]{2022PASJ...74..247A}. Roughly 10\% of the LSST observing time will be used in mini-surveys and the six DDFs. Through a combination of depth and cadence, the DDFs are expected to provide the most complete catalogs of AGNs and their host galaxies \citep{2018arXiv181106542B, 2022ApJS..262...15Z,LSST_Kovacevic2022}.

\begin{sidewaysfigure}
    \centering
    \includegraphics[width=\textwidth, height=0.45\textwidth]{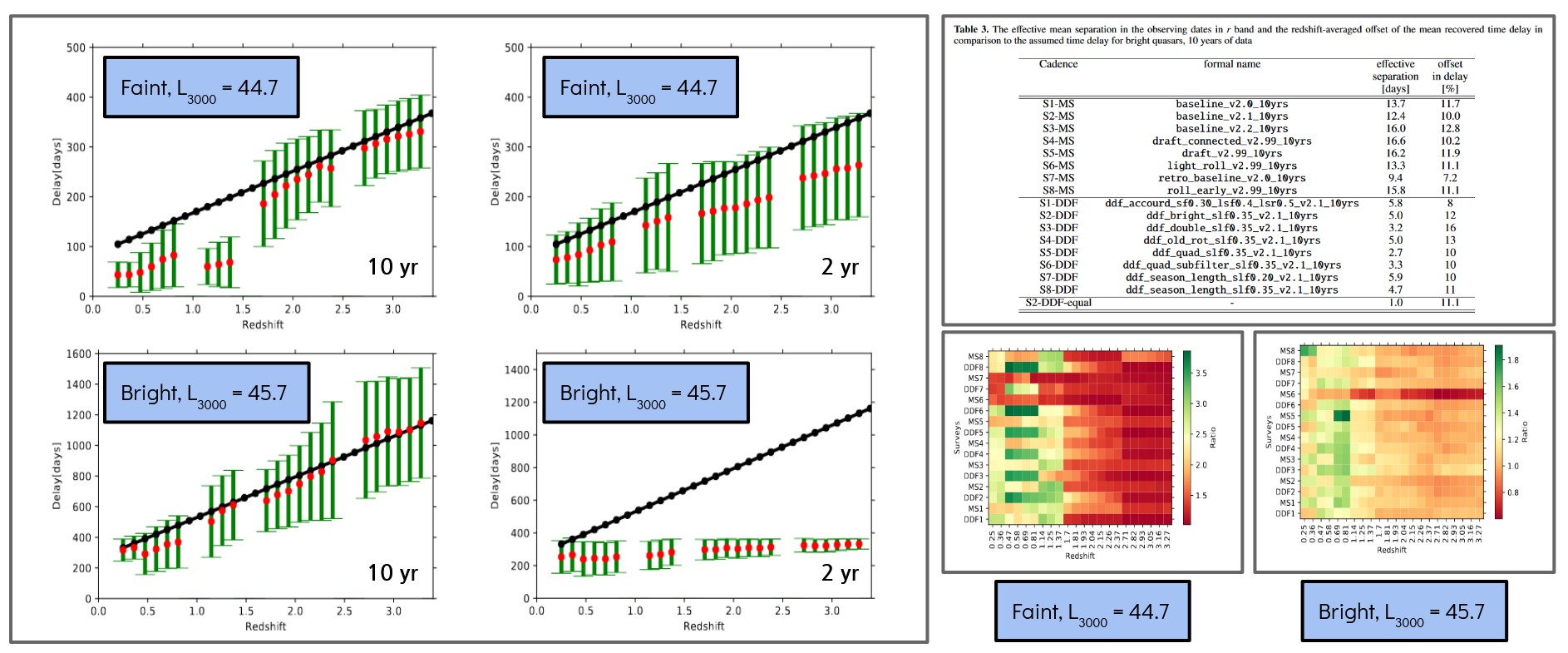}
    \caption{Expectations from the time-lag recovery for AGNs to be observed with LSST: rest-frame BLR time-lag versus AGN luminosity (Clockwise from top left:) In the upper panel, we compare the adopted and recovered time delays as a function of redshift for a faint AGN ($\log L_{3000}$ = 44.7), while in the lower panel, we present the same comparison for a bright AGN ($\log L_{3000}$ = 45.7). These data points are obtained from 10 years of simulated observations in the ELAIS-S1 field, one of the Deep Drilling Fields (DDFs). The green error bars represent the standard deviation expected in a single-source measurement. This is determined by utilizing 100 statistically equivalent simulations; (right panels) Same as the left panel but from the 2 years of simulated observations in the DDFs. The table situated in the top right corner provides information about two key aspects: It showcases the effective mean separation in observing dates in the $r$ band and the redshift-averaged offset of the mean recovered time delay when compared to the assumed time delay for bright quasars. This data is derived from a 10-year simulation, considering various survey strategies proposed by the scientific collaborations within LSST, and in the bottom left and bottom right panels, one can observe the color-coded relative systematic error in determining the time delay. This information pertains to both faint (left panel) and bright (right panel) AGN, and it's based on a 10-year simulation across 8 Main Survey (MS) and 8 DDF cadences. For these simulations, an arbitrary location in the sky, specifically (0, -30), was chosen for the MS, while for DDF, the ELAIS-S1 field centered at (9.45, -44.025) was used as the reference location. Abridged figure based on \citet{Czerny_Panda_etal_2023}. Reproduced with permission from Astronomy \& Astrophysics, © ESO}
    \label{fig:lsst_prm}
\end{sidewaysfigure}

Our modeling approach relies on stochastic light curves, which provide a robust representation of active galactic nuclei (AGN) variability. We utilize the Timmer-K\"onig algorithm to generate synthetic light curves, which are subsequently sampled according to specific cadences. We explore various cadence simulations provided by OpSims from LSST \citep{2018arXiv181200516S,LSST_Ivezic_etal_2019,2022MNRAS.512.5580S,LSST_Kovacevic2022,2022ApJS..258....3R} for the MS and the DDFs as listed in Figure \ref{fig:lsst_prm} (top right panel). For more details on the properties of each of these simulations we refer the readers to \citet{Czerny_Panda_etal_2023}. These simulated light curves are designed to include the expected contamination from prominent emission lines (H$\beta$, Mg II, CIV), the Fe II pseudo-continuum, and stellar emissions. By selecting appropriate photometric bands suitable for the source's redshift, we compare the predicted line time delay with the retrieved delay across 100 statistical realizations of the light curves.

The outcomes representing the entirety of the 10-year monitoring within one of the Deep Drilling Fields (DDFs), specifically ELAIS S1, are displayed in the left panels of Figure~\ref{fig:lsst_prm}\footnote{For the Main Survey (MS), we adopted the sky location (0, -30), while for the DDF, the central coordinates of ELAIS-S1 at (9.45, -44.025) were utilized.}. It's evident that for bright quasars, the results obtained from the DDF exhibit improvements in the smallest redshift range (below 1.0) compared to the MS (refer to Figure 4 in \citet{Czerny_Panda_etal_2023}), wherein more frequent coverage facilitates accurate determination of the time delay, especially in cases where it's relatively short. This difference becomes highly significant for faint quasars. While the MS lacks reliable sampling for them, those situated within the DDFs can be robustly measured for redshifts surpassing 1.8. This becomes intriguing and pivotal, considering that fainter quasars are anticipated to dominate the quasar population, thus the DDFs have the potential to detect numerous faint quasars. Notably, about 80\% of quasars in recent SDSS data releases are brighter than $\log L_{3000} = 44.7$, rendering them well-measurable within DDFs. In contrast, only approximately 15\% of quasars in the MS possess sufficient luminosity to yield adequately measured delay timescales. For additional insights, readers are directed to Figure 3 and Table 3 in \citet{Czerny_Panda_etal_2023}. The decreased accuracy at lower redshifts can be partly attributed to the significant intervals between observing seasons. The noticeable underestimation of the time delay occurs for redshifts between 1.0 and 1.5, primarily because the simulations for the chosen luminosity level yield an expected time delay of roughly 180 days. For redshifts below 0.5, segmenting data into separate seasons could offer improvement.

We also scrutinize the feasibility of yielding informative outcomes from the first two years of LSST data. For the MS, reliable results are not anticipated due to the prevailing sampling. In contrast, the denser cadence in the DDFs renders it possible to measure time delays of the order of a year. Especially when dealing with faint AGN, where anticipated time delays are below 400 days, measurements are viable (see Figure~\ref{fig:lsst_prm}, top panel in the middle). The actual measurements and the expected values agree within a 1$\sigma$ margin of error at all redshifts. However, a systematic mean offset from the expected values across all redshifts is evident. This offset is attributed to the relatively short data sets. Nevertheless, these measurements offer value for specific statistical investigations, even though the presence of a systematic offset of approximately $\sim$40\% should be taken into account. This offset depends on the precise source luminosity, indicating the need for additional simulations to refine the results. In the case of bright AGN, the expected time delays are so extended that meaningful results are only attainable for objects with redshifts below 0.3 when there are only two years of data available. Otherwise, the recovered delay reaches saturation at the code's maximum limit, which is determined as the lesser of half the duration of the data set and twice the expected time delay (see Figure~\ref{fig:lsst_prm}, bottom panel in the middle).

Considering a single year of data usage renders the situation even more challenging. Reliable measurements are not viable for bright sources across any redshift range. For faint sources, only promising results from low redshifts emerge. In both scenarios, AGN with redshifts surpassing 0.7 is inaccessible, which encourages cautious searches that may consider setting a lower redshift limit, approximately around $\sim$0.4. Nonetheless, it is promising that specific AGN emission line time delays can still be measured, even within shortened monitoring periods characterized by highly irregular cadences.

To evaluate the potential cadences suggested for LSST's various survey strategies, we tested eight different scenarios for both the MS and DDFs (see top right panel of Figure~\ref{fig:lsst_prm}). For succinct representation, error bars were omitted as their impact remains relatively stable. Instead, we employed color-coded plots that depict the relative change between the mean derived time delay, $\tau_{\rm derived}$, and the time delay assumed in simulations, $\tau_{\rm adopted}$, expressed as $\delta = \frac{\tau_{\rm adopted}}{\tau_{\rm derived}}$. The dimensionless value exhibits a strong dependence on source luminosity and is depicted as a function of redshift separately for faint and bright AGN (refer to the bottom right panels of Figure~\ref{fig:lsst_prm}). These assessments were performed using a 10-year simulated cadence. Across most cadences, results generally align with those previously obtained. The results for bright AGN are quite gratifying, particularly for moderate to high redshifts. However, there is a recurring issue of underestimating the time delay for AGN located near redshifts of approximately z $\sim$ 0.7 wherein the derived value of is too short. As discussed by \citet{czerny2013}, obtaining meaningful results for bright quasars requires five measurements per year with uniform spectroscopic coverage or twice that number for photometric measurements with non-uniform sampling. Lightcurves providing fewer data points cannot be effectively utilized. For fainter quasars, the majority of cadences anticipate time delays that are too short to be sufficiently covered in the MS mode. In the case of DDFs, the outcomes exhibited a similar qualitative trend, permitting a quantitative comparison through two global parameters computed for each cadence. One parameter indicates the mean time interval between visits, considering observations on the same day as a single exposure. The second parameter signifies the redshift-averaged value of $\delta$. These values for bright quasars are presented in the top right panel of Figure~\ref{fig:lsst_prm}. 

We illustrate that it is feasible to obtain emission line time delays from the photometric data provided by LSST for a significant proportion of quasars. Our findings reveal the following insights: (i) For quasars surpassing $\log L_{3000} = 45.7$, the cadence within the MS generally proves sufficient for measuring line time delays relative to the continuum. Although individual measurement errors in time delays are relatively high, around 40\%, most cadences exhibit a systematic offset of 10-15\%. Hence, aggregating measurements from numerous quasars enables statistical studying trends like the radius-luminosity relation; (ii) For quasars with luminosities dimmer than $\log L_{3000} = 44.7$, it is not advisable to use the MS. Determining a practical redshift limit for usability in the case of intermediate luminosity quasars will require simulations; (iii) Successfully measuring line time delays in quasars with luminosities below $\log L_{3000} = 44.7$ is achievable within the DDFs. In such cases, even the initial two years of data are sufficient, and extending the dataset does not significantly improve delay measurements at lower redshifts unless the data is sampled over shorter intervals or subjected to detrending; (iv) Dense sampling can also aid in studying bright quasars situated within DDF fields, marginally reducing individual errors to below $\sim 30\%$; (v) Each of the cadences investigated introduces a systematic offset of approximately 10\% between the assumed setup delay and the recovered time delay. The offset persists even with numerous measured quasars, necessitating correction via numerical simulations. The offset's magnitude hinges on quasar properties, cadence, photometric errors, and the chosen time delay measurement method, (vi) While certain cadences fare better than others, the cadence does not appear to be a critical determinant for line delay measurements.

Hence, by carefully selecting source luminosities and their associated redshift ranges, we can anticipate reliable time delay measurements for a substantial portion of quasars observed within the Main Survey (MS) upon its completion. The exact fraction can be determined by configuring the observational cadence. Significantly, results for intrinsically fainter sources in Deep Drilling Field (DDF) locations with lower redshifts can be derived from data acquired even during the project's inaugural year. For more comprehensive information on these results and how the choice of survey strategies influences the recovery of time delays in the Broad Line Region (BLR) for the observed AGNs throughout the campaign's duration, we direct readers to the study by \citet{Czerny_Panda_etal_2023}.

\subsection{Accretion disk continuum modeling and contamination from the BLR}
\label{sec:2.2}

The reverberation technique has been employed to measure the size of the accretion disk in AGNs using continuum time delays \citep[see e.g.,][]{2018ApJ...854..107F, 2018ApJ...862..123M, 2019MNRAS.482.2788S, 2020A&A...636A..52C, 2021MNRAS.504.4337V, 2022ApJ...925...29C, 2022ApJ...929...19G, 2022ApJ...940...20G, 2022MNRAS.511.3005J, 2023MNRAS.521..251H, 2023A&A...672A.132F, 2023ApJ...953..137M}. Initially proposed by \citet{collier1999} (with a torus-based variation by \citet{oknyanskij1999}), this approach involves evaluating the size of the accretion disk across different wavelengths and then comparing it to the classical accretion disk model proposed by \citet{ss73}. This comparison is made possible through a particular scaling of monochromatic flux and effective temperature relative to the product of the black hole mass and accretion rate \citep[as discussed in][]{panda18b}, allows for directly determining the source's distance using the measured time delay ($\tau$) between two wavelengths and the observed monochromatic flux ($f_{\nu}$) at one of these wavelengths. This distance estimation remains independent of the black hole mass and accretion rate. Theoretically, the time delay is expected to scale with wavelength $\lambda$ as $\tau \propto \lambda^{4/3}$, and the coefficient of proportionality encompasses observed flux and distance, enabling redshift-independent distance determination.

Empirical monitoring of several sources corroborated the anticipated delay pattern \citep[e.g.][]{collier1999,cackett2007,pozo2019,lobban2020}, specifically the proportionality $\tau \propto \lambda^{4/3}$. Yet, the determined proportionality constant from observations often significantly exceeded the theoretical prediction, varying by roughly 40\% up to several factors, when employing the standard cosmology \citep[e.g.][]{collier1999,cackett2007,lobban2020,GuoW2022,Gonzalez23}. Certain sources/papers achieved alignment with expectations. For instance, a correction for internal AGN reddening is necessary to be consistent with $\nu^{1/3}$ (see e.g. \citet{Weaver_Horne_2022}). Neglecting to incorporate this correction can lead to a notable underestimation of AGN luminosity, such as a tenfold underestimation in the ultraviolet and approximately a fourfold underestimation in the optical (\citet{Gaskell_2017}). This bias in luminosity can consequently lead to underestimated AD sizes as recently shown by \citet{Gaskell_etal_2023} for NGC5548. Conversely, as proposed by \citet{pozo2019}, an underestimation of the black hole mass owing to the unknown geometry of the BLR is suggested as the root of a biased time-delay spectrum (as depicted in Figures 4 and 6 of their study). Models considering the X-ray illumination of the accretion disk (AD) could also explain the observed longer delays in specific cases, taking into consideration factors such as the height of the corona or the extended nature of the reprocessor \citep[][]{kammoun2021_data,Kammoun23}, while others reported disparities at specific wavelength bands near the Balmer edge \citep[e.g.][]{edelson2015, Kammoun2019, McHardy2018,cackett2018,hernandez2020,cackett2020}. Indications arose of substantial contamination from the more distant reprocessing region - the BLR - especially around the Balmer ($\sim$3647\AA) and Paschen ($\sim$8206\AA) edges (e.g., see Figure 1 left panel in \citet{chelouche2019}), collectively referred to as the diffuse continuum emission (DCE, \citep{Korista_Goad_2001,PozoNunez_etal_2023}). In his recent research, \citet{netzer2021} concluded that in the case of most examined sources, reprocessing primarily takes place in the BLR, thereby extending the continuum time delay beyond the anticipated disk value. A similar hypothesis had previously arisen for NGC 5548, as presented by \citet{lawther2018}. Contributions from non-disk components, such as the \feii{} pseudo-continuum and the Balmer continuum, were also considered \citep{wills1985}, also contribute to the contamination, which becomes significant when employing photometric broadband filters, as in the case of LSST. For instance, \cite{netzer2021} recently revealed a new delay-luminosity relationship for specific sources driven solely by the DCE of the BLR, unrelated to the accretion disk itself. This intricate relationship's origin remains subject to debate, making it challenging to discern via observations. Thus, comprehending these effects is vital to evaluate their impact and develop methods to disentangle genuine continuum information from the accretion disk. The time delay of the accretion disk, studied across numerous objects, will aid in untangling these effects, underscoring LSST's pivotal role.

In our recent study (\citet{2023A&A...670A.147J}), we employ numerical simulations to determine if the assumed disk and BLR region geometries can be extracted from time delay measurements. We generate synthetic light curves to replicate incident radiation, taking into account parameters that describe both the accretion disk and the BLR reprocessing. In our modeling, we estimate the expected time delays, incorporating contributions from both the reprocessing of incident X-ray flux by the disk and subsequent reprocessing by the BLR. Our goal is to ascertain whether these two effects can be differentiated. Our findings indicate that the impact of BLR scattering on predicted time delays mirrors the effect of increasing X-ray source elevation without BLR contribution. Consequently, continuum time delays rise linearly with larger BLR contribution, introducing added degeneracy when attempting to deduce system parameters from observed time delays. Nonetheless, both effects alter the slope of the delay-wavelength curve when viewed in logarithmic space. This alteration provides a means to extract the pure disk time delay, which is essential for cosmological applications.

In another recent study (\citet{PozoNunez_etal_2023}), we explore determining the AD sizes of quasars and assess the LSST survey's proposed cadence using photometric reverberation mapping (PRM). Our extensive simulations factor in LSST survey characteristics and quasars' intrinsic properties. These simulations characterize light curves, enabling the determination of AD sizes using various algorithms - we account for contamination by a host galaxy in the photometric aperture using the flux variation gradient method \citep{Choloniewski_1981,Winkler_etal_1992, PozoNunez_etal_2014,Gaskell_etal_2004, Gaskell_2007, Pozonunez12}. We also refer the readers to Figure 2 in \citet{PozoNunez_etal_2023}. We include the nuclear reddening contributions caused by the internal reddening of the AGN \citep{Gaskell_etal_2004, Gaskell_2007}, as well as the contribution from the variable DCE \citep{Korista_Goad_2001, Korista_Goad_2019} and BLR emission lines \citep{VandenBerk_etal_2001, Glikman_etal_2006}. We determine that: (i) Reaching a signal-to-noise ratio (S/N) of a minimum of 100, with BLR emission line contribution staying below 10\% within the bandpasses, facilitates the recovery of time delays with 5\% and 10\% accuracy for 2-day and 5-day time sampling, respectively, for quasars in the range $1.5 < z < 2.0$. Accuracy of 10-20\% is attainable for quasars at $z < 1.5$ only if BLR emission line contribution remains under 5\%. Augmenting S/N does not notably enhance results; (ii) Assuming an optically thick, geometrically thin accretion disk model, the obtained time delay spectrum aligns well with estimated black hole masses, with an accuracy of approximately 30\%. Under the prevailing observational circumstances of LSST, this accuracy range encompasses intermediate black hole masses in the vicinity of $\sim 10^{8} - 10^{9}$ M$_{\odot}$. For the DDFs, with possible 1-2 day time sampling, the range could extend to $\sim 5\times10^{7}$ M$_{\odot}$. We recommend strategies for LSST's observing approach to effectively recover continuum time lags for a substantial number of AGNs, including correcting fluxes in each photometric band for host galaxy contamination and internal reddening, as well as scaling the BLR's variable DCE lag spectrum to account for luminosity differences. It's important to emphasize that our analysis extends beyond LSST and can be applied to any future photometric reverberation mapping survey of accretion disks. Considering cosmological time dilation, studies involving the reverberation mapping of continuum-emission lines in luminous quasars at around $z\sim2$ require approximately 15 years of observations (for example, using CIV lines). Accretion disk time delays are roughly 10 times shorter than those of the BLR (as discussed in \citet{2023ApJ...948L..23W}), which holds the potential to establish a more precise AD size-luminosity relationship. This, in turn, enables quasars to function as effective standard cosmological candles up to high redshifts. Nevertheless, achieving sufficient time sampling and conducting specialized surveys are crucial for the exploration of specific quasar populations. In this context, we are planning a photometric reverberation mapping campaign for accretion disks (AD PRM) targeting quasars up to z $\le$ 1.5 utilizing optical and near-infrared telescopes at the Cerro Armazones Observatory in Chile. Further information about this upcoming survey will be presented in a forthcoming paper.







\section{Concluding remarks}
\label{sec:conclusions}

Through this review, we have highlighted some recent progress in the studies of accretion disk structure and the development of a database of AGN SEDs, especially for the sources accreting at or above the Eddington limit (see Section \ref{sec:2}). We showcase how the increase in the accretion rate needs to be accounted for changes in the accretion disk structure which then gives rise to the additional anisotropic emission. At close to the Eddington accretion rates, the accretion disk transitions towards a ``slim'' disk scenario wherein we see the formation of a \textit{funnel-like} feature at the innermost region closest to the SMBH. Such a change in the disk structure could explain the dichotomy of the incident flux received by the high- and low-ionization line-emitting regions. The low-ionization lines located closer to the disk's surface end up receiving a much weaker, filtered continuum which in turn allows for these clouds to move in closer to the central source (see Figure \ref{fig:qms_disks}) - this would clarify why the high-accreting sources tend to show shorter BLR lags in the \rblr{}-\lopt{} relation. We additionally derive the modeled SEDs which are dependent on the accretion rates (going from sub-Eddington to super-Eddington rates) and also highlight how the continuum emission varies as a function of the viewing angle (see Figure \ref{fig:sed-models_slimdisks}).

We then discuss the effectiveness of the BLR radius vs. AGN luminosity relation(s) in providing us with an independent way to estimate the luminosity distances to these cosmic sources, help build the Hubble diagram, and eventually test and retrieve cosmological parameters of our Universe (see Section \ref{sec:4}). Stepping back to consider the broader perspective, we can build the Hubble diagram utilizing the corrected luminosity distances from Eq. \ref{eq:dl}, adjusted following Eq. \ref{eq:rcorr}, along with the corresponding redshift values for each source. The crux of this approach lies in possessing measurements of time delays (e.g., derived from Eq. \ref{eq:rcorr}) and the monochromatic flux of the AGN derived from spectroscopy. This enables the estimation of luminosity distances independent of accretion properties. Consequently, extensive sets of AGNs subject to reverberation mapping can be harnessed as effective cosmological distance indicators \citep[][]{collier1999, elvis2002, horne2003, Haas2011, panda_2019_frontiers}. This inclusive approach opens the door to exploring the redshift-dependent cosmological parameters' evolution, thereby aiding in reconciling the Hubble tension - the discrepancy between the estimated Hubble constant values in the late and early Universe (we refer the readers to the right panel in Figure \ref{fig:quasars_cosmo}). The ideas and findings presented in this study hold immense potential, setting the foundational platforms for the extensive array of AGNs which will be investigated using ongoing and future ground-based telescopes with 10-meter-class capabilities \citep[e.g., Maunakea Spectroscopic Explorer,][]{2019BAAS...51g.126M} and 40 meter-class capabilities \citep[e.g., The Extremely Large Telescope,][]{2015arXiv150104726E}; in addition to space-based missions such as the JWST \citep{2006SSRv..123..485G, 2022A&A...661A..80J} and the Roman Space Telescope \citep{2013arXiv1305.5422S}. It is crucial to have access to a growing database of high-quality data, including photometric, spectroscopic, and interferometric information spanning higher redshifts which are essential for advancing our ever-expanding theoretical understanding of the mechanisms and evolution of AGNs and for utilizing them as tools for estimating cosmic distances.

We then demonstrate a novel BH mass scaling relation traced by the intensities of high-ionization, forbidden emission lines (known as coronal lines) that we have discovered (see Section \ref{sec:3}, also Figure \ref{fig:coronal-lines}). These coronal lines, a bulk of them produced in the NLR, are shown to evolve with the change in the underlying accretion disk structure, which is the primary source of the photons that produce emission of such lines (see Section \ref{sec:3.2}). A recent intriguing avenue that has gained prominence is the potential of utilizing high-ionization emission lines within the optical and near-infrared spectral ranges as effective indicators for accurately characterizing the dynamics and physical properties of gas ionized by the AGN, and further associated with X-ray emission lines \citep{2022MNRAS.511.1420T}. These emission lines, often referred to as ``footprint lines'', emanate from gas within a comparable range of ionization states as those of oxygen and neon ions configured in H- and He-like states - these same ions are responsible for generating X-ray emission lines. Instruments such as the Hubble Space Telescope/STIS and JWST/NIRSpec, which operate in the optical and infrared regimes, possess the capacity to detect these footprint lines. Consequently, these lines present a promising opportunity to glean insights into the kinematics of the extended gas responsible for emitting X-rays. Building upon prior foundational research \citep{Prieto_etal_2022, 2022MNRAS.511.1420T}, a natural progression is to establish an extensive database encompassing properties of coronal lines across a diverse spectrum of active galaxies. Such a compilation would prove invaluable in identifying new AGNs, particularly in higher redshift domains, a potential magnified by the capabilities of the forthcoming JWST data. An especially noteworthy application of these coronal lines' diagnostics is their potential to detect and unveil accreting black holes within the intermediate mass range, known as Intermediate Mass Black Holes (IMBHs). As the JWST builds on its recent success \citep{2022ApJ...926..161B, 2022ApJ...940L...5U, 2023ApJS..265....5H, 2023A&A...677A..88B, 2023ApJ...942L..17O, 2023ApJ...942L..37A, 2023ApJ...953L..29L, 2023ApJ...954L...4K, 2023ApJ...955L..24G, 2023A&A...672A.128C, 2023A&A...677A.145U, 2023ApJ...948..112C, 2023ApJ...953...10C, 2023arXiv230311946H, 2023arXiv230605448M, 2023arXiv230607320L, 2023arXiv230905714G}, we find ourselves in an opportune position to validate the theoretical predictions of coronal line emissions and their fundamental linkage with central accreting black holes.

We showcased our efforts in preparation for the upcoming Vera Rubin Observatory's Legacy Survey for Space and Time (LSST) in the context of the accretion disk and BLR time delay measurements using photometric monitoring and scrutinizing the survey strategies that are under discussion (see Sections \ref{sec:2.3} and \ref{sec:2.2}). We critically assess the various survey strategies for the LSST by accounting for them in our time-lag recovery pipeline as part of our in-kind contribution to LSST. We test the reliability of the time-lag measurements for multiple broad emission lines across a wide redshift and luminosity range and include the contribution from the host galaxy and other contaminants, such as the \feii{} pseudocontinuum (see Figure \ref{fig:lsst_prm}). Through our analyses, we find that the success rate of lag-recovery is significantly higher in the Deep Drilling Fields (DDFs) where the dense sampling is beneficial to reduce the uncertainties in the lag measurements \citep{Czerny_Panda_etal_2023}. In a corollary work, we explore determining the accretion disk (AD) sizes of quasars and assess the LSST survey's proposed cadence using photometric reverberation mapping. Our extensive simulations factor in LSST survey characteristics and quasars' intrinsic properties. Similar to the previous work, we account for the contamination from the host galaxy, including nuclear reddening contributions caused by the internal reddening of the AGN, as well as the contribution from the variable diffuse continuum emission (DCE) and BLR emission lines. We find from our numerical simulations, that reliable time-lags for the accretion disk can be obtained from the LSST for a few thousands of quasars, specifically in the DDFs with a possible 1-2 day time sampling. We test also how the contribution from the BLR lines can constrain the redshift windows which should be prioritized to recover accurate AD time lags \citep{PozoNunez_etal_2023}. Reverberation mapping of high-z quasars (z$\sim$2) requires approximately 10-15 years of observations. Accretion disk time delays are roughly 10 times shorter than those of the BLR, which holds the potential to establish a more precise AD size-luminosity relationship. This, in turn, enables quasars to function as effective standard cosmological candles up to high redshifts.

\vspace{6pt} 


\authorcontributions{We thank the anonymous reviewers whose suggestions helped us improve the overall presentation of the results and findings in this review. The idea for the article and manuscript preparation was made by SP. PM, BC, ARA, and FPN assisted with the writing in different sections presented in the manuscript. Each of the co-authors helped with the review and editing of the presented version. All authors have read and agreed to the published version of the manuscript.}

\funding{This research received no external funding.}

\dataavailability{Data used in this work can be provided upon request to the corresponding author.} 

\acknowledgments{SP acknowledges the Conselho Nacional de Desenvolvimento Científico e Tecnológico (CNPq) Fellowships (164753/2020-6 and 313497/2022-2). BC and FPN acknowledge funding from the European Research Council (ERC) under the European Union’s Horizon 2020 research and innovation program (grant agreement No 951549). FPN gratefully acknowledges the generous and invaluable support of the Klaus Tschira Foundation. SP is grateful to the organizers of the 14th Serbian Conference on Spectral Line Shapes in Astrophysics, Bajina Bašta, Serbia, June 19 - 23, 2023, where the work was presented in the form of an invited talk.}

\conflictsofinterest{The authors declare no conflict of interest.} 


\abbreviations{Abbreviations}{
The following abbreviations are used in this manuscript:\\

\noindent 
\begin{tabular}{@{}ll}
AD & Accretion disk\\
AGN & Active galactic nuclei\\
BH & Black hole\\
BLR & Broad-line region\\
CL & Coronal lines\\
DCE & Diffuse Continuum Emission\\
DDF & Deep Drilling Field\\
DL & Luminosity distance\\
ECDFS & Extended Chandra Deep Field-South\\
EDFS & Euclid Deep Field Survey\\
FWHM & Full width at half maximum\\
GNIRS & Gemini Near-Infrared Spectrograph\\
IMBH & Intermediary mass black hole\\
IP & Ionization potential\\
ISCO & Innermost stable circular orbit\\
LSST & Legacy Survey of Space \& Time\\
MS & Main Survey\\
NLR & Narrow line region\\
NLS1 & Narrow-Line Seyfert 1\\
NIR & Near infrared\\
PRM & Photometric Reverberation Mapping\\
RM & Reverberation Mapping\\
SDSS & Sloan Digital Sky Survey\\
SED & Spectral energy distribution\\
SEAMBH & Super-Eddington Accreting Massive Black Holes\\
SMBH & Supermassive black hole\\
SNe & Supernovae\\
SS & Shakura \& Sunyaev\\
VLTI & Very Large Telescope Interferometer\\
WC & Warm Corona
\end{tabular}
}

\begin{adjustwidth}{-\extralength}{0cm}
\reftitle{References}

\bibliography{reference}
\PublishersNote{}
\end{adjustwidth}
\end{document}